\documentclass[aps,physrev,reprint,superscriptaddress,fleqn,floatfix]{revtex4-2}

\usepackage{amsmath, bm, mathtools, color}
\usepackage{graphicx} 
\usepackage{hyperref}
\usepackage{siunitx}
\DeclareSIUnit{\litre}{l}

\allowdisplaybreaks
\allowdisplaybreaks
\usepackage{xr} 
\usepackage[normalem]{ulem}
\usepackage{xcolor}

\DeclareMathOperator*{\argmax}{argmax} 

\begin{document}
\suppressfloats[t]


\title{Bayesian comparison of Langevin dynamics for cell motility from positional observation}


\author{Yusuke Kato}
\email{yusukeka@umich.edu}
\affiliation{Gilbert S.~Omenn Department of Computational Medicine and Bioinformatics, University of Michigan, Ann Arbor, 48109-2218, MI, USA}
\affiliation{Department of Complexity Science and Engineering, Graduate School of Frontier Sciences, The University of Tokyo, Kashiwa, Chiba 277-8561, Japan}

\author{Jan Albrecht}
\affiliation{Institute of Physics and Astronomy, University of Potsdam, 14476 Potsdam, Germany}

\author{Ted Moldenhawer}
\affiliation{Institute of Physics and Astronomy, University of Potsdam, 14476 Potsdam, Germany}

\author{Robert Gro\ss mann}
\affiliation{Institute of Physics and Astronomy, University of Potsdam, 14476 Potsdam, Germany}

\author{Carsten Beta}
\affiliation{Institute of Physics and Astronomy, University of Potsdam, 14476 Potsdam, Germany}
\affiliation{Nano Life Science Institute (WPI-NanoLSI), Kanazawa University, Kakuma-machi, Kanazawa 920-1192, Japan}



\date{\today}

\begin{abstract}
We develop a Bayesian framework for model comparison of second-order Langevin dynamics from position-only trajectories. While approximate increment likelihoods for nonlinear position-only inference have been formulated previously, a unified evidence-based framework for comparing multiple second-order models under positional observation has remained lacking. Here we address this problem by combining exact increment likelihoods for linear Gaussian models with a previously proposed approximate likelihood for nonlinear dynamics. 
Synthetic-data benchmarks show reliable recovery of the generating model at fine sampling intervals and progressive loss of identifiability under coarse temporal sampling. Application to {\it Dictyostelium discoideum} trajectories demonstrates that the statistically supported model depends strongly on temporal resolution. Moreover, the selected models reproduce key statistical properties of the experimental trajectories, providing additional support for the model-comparison results. Our framework therefore offers a practical approach to evidence-based comparison of partially observed stochastic dynamics.
\end{abstract}


\maketitle

\section{Introduction}

Random motion is ubiquitously observed in a variety of physical, chemical, and biological systems~\cite{codling2008random,metzler2014anomalous,bechinger2016active}. A prominent biological example is cell motility~\cite{selmeczi2008cell}, the spontaneous movement of cells, which occurs across a wide range of cell types, including cancer cells, leukocytes, and keratinocytes~\cite{aranson2016physical}. Beyond its fundamental biophysical interest, cell motility has important biomedical implications, including the regulation of cancer and tumor-reactive immune-cell migration for cancer treatment~\cite{stuelten2018cell} and the targeted delivery of anti-inflammatory substances by macrophages to promote early removal of brain hematomas~\cite{kapate2024backpack}. 

To characterize apparently random cell movement and elucidate its underlying mechanisms from mathematical and physical perspectives, numerous models have been developed from experimentally observed trajectories~\cite{klages2024cell}.  The mean squared displacement (MSD), which characterizes the expected displacement over different time lags, typically shows ballistic motion at short lags and normal diffusion at long lags~\cite{bruckner2024learning}. This behavior motivates the classical persistent random motion model~\cite{gail1970locomotion}, in which the velocity follows an Ornstein--Uhlenbeck (OU) process~\cite{uhlenbeck1930theory}, and which still serves as a reference model for motility~\cite{bruckner2024learning}. The velocity autocorrelation function of several cell types exhibits double-exponential decay, motivating velocity dynamics with exponential memory-kernel terms~\cite{selmeczi2005cell, selmeczi2008cell, takagi2008functional, mitterwallner2020non} and more complex stochastic models~\cite{campos2010persistent,li2011dicty}. Beyond temporal correlation statistics, the shape of the velocity distribution provides another basis for model construction. In particular, the ring-shaped velocity distribution reported in Ref.\cite{li2008persistent} is consistent with the stationary distribution of a preferred-speed model\cite{erdmann2000brownian,romanczuk2012active}, which we refer to here as the Mexican-hat model [Eq.~\eqref{eq:Mhat}].

In parallel with physically motivated modeling, data-driven inference frameworks for stochastic dynamics have been developed extensively~\cite{boninsegna2018sparse,bruckner2020inferring,frishman2020learning,tabar2024revealing}, including applications to cell motility~\cite{bruckner2019stochastic,mitterwallner2020non,bruckner2024learning}. Some of these approaches formulate model selection as the statistical identification of the most suitable dynamics from a prescribed set of basis functions~\cite{gerardos2025principled,kashiwamura2026uncertainty}. Related data-driven comparison methods have also been used to rank candidate models with distinct physical characteristics in a variety of stochastic systems, including collective cell-density dynamics~\cite{ferguson2016inference}, anomalously diffusing microparticles in mucus~\cite{lysy2016model} and mucin hydrogels~\cite{thapa2018bayesian}, and synthetic trajectories of turbulent~\cite{brolly2022bayesian}.

Despite the accumulation of phenomenological models and the progress in data-driven inference, systematic model comparison for cell motility from single-particle tracking data remains limited. A central challenge is to infer second-order dynamics from positional observation. Motility models are often intrinsically described as second-order Langevin models, involving both position and velocity variables~\cite{bruckner2020inferring}, whereas experimental trajectories usually provide only positions. The velocity therefore cannot be observed, and naive velocity reconstruction from positional increments can lead to biased inference~\cite{albrecht2024inferring}.

This challenge has not been covered by many existing data-driven comparison methods, which assume that all state variables can be observed~\cite{gerardos2025principled,kashiwamura2026uncertainty}. A Bayesian comparison of first-order models~\cite{thapa2018bayesian} avoids this issue because the dynamics contain no hidden velocity variable. Conversely, model comparison for second-order dynamics from positional observations has so far been developed mainly for linear Gaussian models, where the full likelihood is analytically tractable~\cite{lysy2016model,brolly2022bayesian}. Thus, a unified framework for comparing general second-order Langevin models from position-only observations is still lacking. 

In the present study, we develop a Bayesian method for comparing linear and non-linear second-order models from position-only trajectory data. Rather than reconstructing velocities, we work with the non-Markovian increment process obtained after integrating out latent velocity variables. This formulation enables different model classes to be treated within a common Bayesian pipeline: for linear Gaussian models, it gives exact increment likelihoods with Toeplitz covariance matrices, whereas for nonlinear models we use a previously proposed approximate likelihood construction~\cite{albrecht2024inferring}. We test the approach on synthetic trajectories and apply it to {\it Dictyostelium discoideum} ({\it D.~discoideum}) motility data, examining whether the selected models recover key trajectory statistics and how model support changes with temporal resolution.

This study has two main contributions. First, we propose a Bayesian model-comparison framework for second-order dynamics from positional data, and validate its efficacy using both synthetic trajectories and experimental data. Although we focus on cell motility, the framework should also be useful for studying other stochastic dynamics under positional observation. Second, applying the framework to cell-motility data shows that the supported model changes with the sampling interval. These results demonstrate that temporal resolution must be considered when interpreting model selection from trajectory data.

This paper is organized as follows. In Sec.~\ref{sec:model}, we briefly introduce our four candidate models. In Sec.~\ref{sec:methods}, we describe the model-comparison procedure: after reviewing Bayesian model comparison, we summarize likelihoods from positional increments for the candidate models and describe the prior distributions for inference. In Sec.~\ref{sec:bench}, we benchmark the method using numerically generated trajectories and examine how temporal coarse-graining affects model identifiability. In Sec.~\ref{sec:experiment}, we apply the method to experimental {\it D.~discoideum} trajectories. We first perform a preliminary analysis of single-cell tracks, then select the most supported model for each trajectory, examine how temporal resolution affects model selection, and recover key statistical properties from selected models. Finally, Sec.~\ref{sec:dis} presents the discussion and conclusions. 

\section{Candidate models}
\label{sec:model}
We consider four candidate models: Brownian motion (BM), the integrated Ornstein-Uhlenbeck process (OU), the Mexican-hat model (Hat), and the memory-kernel model (MK). These models represent overdamped diffusion, persistent random motion, nonlinear velocity-regulated active motion, and memory-driven dynamics, respectively. Below, we briefly review each candidate models. A detailed explanation for the characteristics of each second-order model (e.g., MSD and VACF) is provided in Appendix~\ref{sec:app_model_detail}. 

Throughout, ${\bm r} = (r^{(x)},r^{(y)})^{\top}$ denotes the position in two-dimensional space, and ${\bm v}= (v^{(x)},v^{(y)})^{\top}$ denotes the velocity when a second-order description is used. The driving noise is denoted by ${\bm \xi}(t) = (\xi^{(x)},\xi^{(y)})^{\top}$ and is taken to be normalized Gaussian white noise with component-wise correlations $\langle \xi^{(\alpha)}(t)\xi^{(\beta)}(t')\rangle=\delta_{\alpha\beta}\delta(t-t')$. 

\subsection{Brownian motion}
The first candidate model is the Brownian motion model:
\begin{equation}
    \dot {\bm r} = \frac{d{\bm r}}{dt} = \sqrt{2D} {\bm \xi}(t), 
    \label{eq:BM}
\end{equation}
where $D > 0$ denotes the diffusion coefficient. The BM model is characterized by a MSD that grows linearly in time at every scale [Eq.~\eqref{eq:BM_msd}]. 
This first-order model is include as a null model of effective diffusion and as a baseline against which the additional structure captured by the second-order models can be assessed.

\subsection{Integrated Ornstein-Uhlenbeck process}
As the simplest linear Gaussian second-order model, we next consider the integrated Ornstein--Uhlenbeck (OU) process, in which the velocity itself follows an OU process:
\begin{subequations}
    \label{eq:OU}
    \begin{align}
        \dot {\bm r} &= {\bm v}, \\
        \dot {\bm v} &= - \gamma \bm v + \sqrt{2D} {\bm \xi}(t), \label{eq:OU_v}
    \end{align}
\end{subequations}
where $\gamma > 0$ denotes the linear damping rate and $D > 0$ is the velocity-space noise strength. This model provides a minimal description of persistent random motion~\cite{romanczuk2012active, bruckner2024learning}: the velocity remains correlated over a finite persistence time ($\simeq \gamma^{-1}$) before the position dynamics crosses over to effective diffusion. 

\subsection{Mexican-hat model}
Although the OU model provides a minimal description of persistent random motion, its linear Gaussian dynamics cannot capture empirical nonlinear velocity statistics observed in cell motility data~\cite{li2008persistent}. To represent nonlinear active motions with a preferred velocity, we consider an second-order model with nonlinear velocity-dependent friction~\cite{erdmann2000brownian,romanczuk2012active},
\begin{subequations}
    \label{eq:Mhat}
    \begin{align}
        \dot {\bm r} &= {\bm v}, \\
        \dot {\bm v} &= -\gamma \bm v(\| \bm v \|^2 - v_r^2) + \sqrt{2D} {\bm \xi}(t), \label{eq:intMhat_v}
    \end{align}
\end{subequations}
where $\gamma>0$ is the damping coefficient, $D>0$ is the velocity-space noise strength, and $v_r > 0$ denotes the characteristic velocity of the system. 
Since the velocity-dependent friction possesses the characteristic velocity $\|\bm v\|=v_r$ and thus generates a ring-shaped stationary velocity distribution [Eq.~\eqref{eq:Mhat_stat_dist}], we call this model as the Mexican-hat model. 

\subsection{Memory kernel model}
Finally, we consider a linear Gaussian model in which the velocity dynamics depends on an exponentially weighted history of the velocity:
\begin{subequations}
    \label{eq:fly}
    \begin{align}
        \dot{\bm r} &= {\bm v}, \\
        \dot{\bm v} &= -\gamma \bm v + A  \int_{-\infty}^{t} dt' e^{-a (t-t')} \boldsymbol{v}(t') + \sqrt{2D} {\bm \xi}(t),  \label{eq:fly_v} 
    \end{align}
\end{subequations}
where $\gamma>0$ is the damping coefficient, $D>0$ is the velocity-space noise strength, $A>0$ is the strength of the memory feedback, and $a>0$ is the decay rate of the memory kernel. 
We impose 
\begin{equation}
    \label{eq:fly_assump}
    a \gamma > A, 
\end{equation}
as the stability condition, which is detailed in Appendix~\ref{app:der_memory}, along with its MSD and VACF. 

This model is motivated by the observations that the VACF of cell-motility is better described by a double-exponential function~\cite{selmeczi2005cell, selmeczi2008cell, takagi2008functional} or by exponentially damped oscillations~\cite{li2011dicty} than by a single exponential decay. 
Although these studies use multiplicative-noise models~\cite{selmeczi2005cell, selmeczi2008cell, takagi2008functional} or more detailed descriptions~\cite{li2011dicty}, we instead use a simpler additive-noise model with an exponential memory kernel to preserve analytical tractability. Since the model is a linear Gaussian system as shown in Eq.~\eqref{eq:fly_2}, we can derive an exact Gaussian likelihood for the position increments.

\section{Model comparison from position-only trajectories}
\label{sec:methods}
In experimental settings, trajectories are often observed only as positions sampled at discrete times, while the velocities remain latent. Our goal is to compare and rank the four candidate models introduced above using only these positional observations. Since three of the models are formulated as second-order dynamics with explicit position and velocity variables, their position-only likelihoods must account for the unobserved velocities, either by integrating them out exactly or by using an approximation. 


Below, we let $\mathcal{D}~\coloneqq~\{t_i, \bm r_i\}_{i=0}^{M}$ denote the observed positions $\bm r_i$ sampled at equally spaced times $t_i \coloneqq i \Delta t$, where $M+1$ is the number of data points and $\Delta t$ is the sampling interval. The total observation time is $T \coloneqq M \Delta t$. We denote the positional increment as $\Delta {\bm r}_i \coloneqq {\bm r}_{i+1} - {\bm r}_i$.


\subsection{Bayesian model comparison}
\label{sec:method_bayes}
For a candidate model $\mathcal{M}_i$ with parameter vector ${\bm \theta}_i$, Bayesian inference assigns a posterior distribution to the parameters after observing the data $\mathcal{D}$. By Bayes' theorem,
\begin{align}
    p({\bm \theta}_i|\mathcal{D}, \mathcal{M}_i)
    &=
    \frac{p(\mathcal{D}|{\bm \theta}_i, \mathcal{M}_i)\, p({\bm \theta}_i|\mathcal{M}_i)}{Z_i}, 
\end{align}
where $p(\mathcal{D}|{\bm \theta}_i, \mathcal{M}_i)$ is the likelihood, $p({\bm \theta}_i|\mathcal{M}_i)$ is the prior distribution, and
\begin{align}
    \label{eq:ml}
    Z_i
    &\coloneqq
    \int d{\bm \theta}_i\,
    p(\mathcal{D}|{\bm \theta}_i, \mathcal{M}_i)\, p({\bm \theta}_i|\mathcal{M}_i)= p(\mathcal{D}|\mathcal{M}_i), 
\end{align}
is the marginal likelihood, also called the model evidence. 
If all candidate models have equal prior probabilities, $p(\mathcal{M}_i)=p(\mathcal{M}_j)$, then the posterior probability of a model is proportional to its evidence~\cite{bishop_pattern_2006}:
\begin{equation}
    p(\mathcal{M}_i | \mathcal{D}) \propto Z_i.
\end{equation}
A central challenge for model comparison is the evaluation of the intergral of the marginal likelihood in Eq.~\eqref{eq:ml}. We evaluate this integral numerically using an importance sampling approach;
 details of the numerical procedure are provided in Appendix~\ref{sec:app_method_detail}.

The log Bayes factor between two models is the logarithm of the posterior model-probability ratio. Under equal model priors, it is given by
\begin{equation}
    \label{eq:lbf}
    \ln \left[ \frac{p(\mathcal{M}_i | \mathcal{D})}{p(\mathcal{M}_j | \mathcal{D})} \right] = \ln Z_i - \ln Z_j.
\end{equation}
Thus, a positive value of log Bayes factor indicates support for model $\mathcal{M}_i$ over model $\mathcal{M}_j$. We report log Bayes factors as the quantitative measure of model comparison in the results below.

The Bayesian framework also provides parameter estimates within each candidate model. In particular, we use the maximum a posteriori (MAP) estimator
\begin{equation}
    \label{eq:map}
    \hat {{\bm \theta}}_i \coloneqq \argmax_{{\bm \theta}_i} \, p({\bm \theta}_i|\mathcal{D}, \mathcal{M}_i),
\end{equation}
given as the parameter value that maximizes the posterior distribution. 

\subsection{Likelihood from positional increments}
\label{sec:method_likeli}

Since the BM model is first-order and without hidden variables, it describes the observed positional data as a Markov process. Hence, the likelihood can be written as a product of transition densities leading to a likelihood of the form 
\begin{gather}
     p(\mathcal{D}|{\bm \theta}) = p({\bm r}_0|D) \prod_{i=0}^{M-1} p({\bm r}_{i+1}|{\bm r}_{i}, D)\,.
\end{gather}
Here and below, we suppress the model label $\mathcal{M}_i$ and the model index on $\bm{\theta}$, when the model under consideration is clear. 
The BM model has Gaussian transition densities
\begin{gather}
    \label{eq:BM_likeli}
    p({\bm r}_{i+1}|{\bm r}_{i}, D) = \prod_{\alpha = x,y} \mathcal{N}(\Delta r_i^{(\alpha)};0, 2D\Delta t)\,,
\end{gather}
where $\mathcal{N}(\bm x;\bm \mu, \Sigma)$ denotes a normal distribution in $\bm x$ with mean $\bm \mu$ and covariance matrix $\Sigma$.
For a fixed initial position, the likelihood therefore takes the form
\begin{gather}
    p(\mathcal{D}|\bm{\theta}) = \prod_{\alpha = x,y} \mathcal{N}(\Delta {\bm r}^{(\alpha)}; {\bm 0}, 2 D\Delta t\, I )\,,
\end{gather}
where $I$ is the identity matrix and $\Delta \bm r^{(\alpha)}$ denotes the increment vector 
\begin{equation}
    \label{eq:increment_vec}
    \Delta \bm r^{(\alpha)} \coloneqq (\Delta r^{(\alpha)}_0, \ldots, \Delta r^{(\alpha)}_{M-1})^{\top},
\end{equation}
for each Cartesian component $\alpha \in \{x,y\}$. 

In contrast, for the second-order models, the observed position positions are not Markovian due to the hidden velocity variable. Simply replacing the unobserved instantaneous velocities by the secant velocity
\begin{equation}
    \bm V_i \coloneqq \frac{\Delta {\bm r}_i}{\Delta t},
    \label{eq:secant-velocity}
\end{equation}
for $i=0, \ldots, M-1$, leads to biased estimation, as discussed in Appendix ~\ref{sec:app_lim_secant}. 
Hence, it is necessary to either exactly or approximately integrate out the unobserved degrees of freedom to obtain the likelihood with respect to the observed positions. 

The linearity of the OU and memory kernel models makes them Gaussian processes. This means that the observed positional increments have a jointly Gaussian probability density:
\begin{equation}
    \label{eq:likeli_lin_gauss}
    p(\mathcal{D}|\bm \theta) = \prod_{\alpha = x,y} \mathcal{N}(\Delta \bm r^{(\alpha)}; \bm 0, \Sigma)\,.
\end{equation}
While the covariance matrix of the BM model above was diagonal, the non-Markovianity of the second-order processes leads to dense matrices $\Sigma$ in these cases.
The covariance matrix can be calculated analytically for both the OU model and the memory kernel model. The exact expressions and their derivations can be found in Appendix~\ref{sec:method_likeli_linear}. 

Due to the nonlinear velocity-dependent friction of the Mexican-hat model, exact calculation of its likelihood is unfeasible. We therefore need to rely on approximations. Here, we use the transformed-Gaussian approximation proposed in Ref.~\cite{albrecht2024inferring}, which approximates a non-linear transformation of the positional data as a Gaussian process. Details can be found in Appendix~\ref{sec:method_likeli_nonlinear}.

\subsection{Prior distribution}
\label{sec:method_prior}

\begin{table} 
\caption{Likelihood, estimated parameters, and their priors for each candidate model}
\begin{ruledtabular}
\begin{tabular}{l|lll}
\label{tab:prior}
Model & Likelihood & Parameter & Prior\\
\hline
BM & Eq.~\eqref{eq:BM_likeli} & $D$ & $\mathcal{L}(10^{-4},10^2)$ \\
\hline
OU & Eqs.~\eqref{eq:likeli_lin_gauss} \& \eqref{eq:cov_OU},  & $D, \gamma$ & $\mathcal{L}(10^{-4},10^2)$ \\
\hline
Hat & Eqs.~\eqref{eq:likeli_nonlin_secant} \& \eqref{eq:likeli_nonlin_approx} & $D, \gamma$ & $\mathcal{L}(10^{-4},10^2)$ \\
& & $v_r$ & $\mathcal{L}(10^{-2},10^2)$ \\
\hline
MK & Eqs.~\eqref{eq:likeli_lin_gauss} \& \eqref{eq:cov_MK} & $D, d_1$ & $\mathcal{L}(10^{-4},10^2)$ \\
 & & $u$ & $\mathcal{L}(2,10^3)$ \\
 & & $s$ & $\mathcal{L}(0.02,50)$ \\
\end{tabular}
\end{ruledtabular}
\end{table}

Finally, we describe the parameters estimated for each model and their prior distributions. For all models, we assume independent priors over the components of $\bm\theta=(\theta_1,\ldots,\theta_\nu)^{\top}$, i.e.,
\begin{equation}
    p(\bm \theta) = \prod_{k=1}^{\nu}p(\theta_k). 
\end{equation}

For the BM, OU, and Mexican-hat models, we estimate the parameters that appear directly in the model equations, all of which are positive dimensional parameters. Motivated by the scale-invariant prior, $p(\theta) \propto 1/\theta$~\cite{von2011bayesian, jeffreys1998theory}, we assign a log-uniform prior on $x\in[\alpha,\beta]$:
\begin{equation}
\label{eq:loguni}
    \mathcal{L}(x;\alpha,\beta) \coloneqq \frac{1}{(\ln \beta - \ln \alpha)x}. 
\end{equation}
When the random variable is clear from context, we write this distribution simply as $\mathcal{L}(\alpha, \beta)$. 
This prior assigns equal probability to equal intervals on the logarithmic scale, giving comparable prior weight to different orders of magnitude. This is useful when plausible parameter scales vary across datasets and also makes the prior invariant to changes of units.

For the memory kernel model, independent priors on the original parameters $(\gamma,a,A,D)$ are inconvenient because the admissible range of $A$ depends on $a$ and $\gamma$ through the stability condition~\eqref{eq:fly_assump}. Such priors can also lead to numerical instability in evaluating the Toeplitz covariance~\eqref{eq:cov_MK} when the two characteristic decay rates are both very small. Based on the reparametrization in terms of $(d_1, d_2, C_1,C_2)$ given in Eq.~\eqref{eq:def_diBi}, we therefore parametrize the model in terms of the smaller decay rate $d_1$, the ratio of the two decay rates $u=d_2/d_1$, the ratio of the corresponding VACF amplitudes $s=C_2/C_1$, and the noise intensity $D$.
Here, $d_1$ and $d_2$ characterize the two relaxation time scales, whereas $C_1$ and $C_2$ determine the relative contributions of the two exponential components to the VACF in Eq.~\eqref{eq:fly_VACF}. We assign independent priors to $(d_1,u,s,D)$. Following the same scale-invariance argument used above, we use log-uniform priors for the dimensional scale parameters $d_1$ and $D$. Although $u$ and $s$ are dimensionless, they are ratios of positive quantities, and we therefore also assign them log-uniform priors.

The supports of the log-uniform priors, i.e., the values of $\alpha$ and $\beta$ in Eq.~\eqref{eq:loguni}, are summarized in Table~\ref{tab:prior}. For most parameters, we use broad priors to cover plausible values spanning several orders of magnitude. Since $v_r$ in the Mexican-hat model and $u$ and $s$ in the memory kernel model distinguish these models from the alternatives, we use the narrower prior to enhance model identifiability. In particular, we set the lower bound for $u$ to $2$ to maintain a clear separation between the memory kernel and OU models: When $d_1 \simeq d_2$, corresponding to $u\simeq1$, one has $A\simeq0$, and the memory kernel model~\eqref{eq:fly} approaches the OU model~\eqref{eq:OU}.

The marginal likelihood~\eqref{eq:ml} depends on the choice of priors~\cite{llorente2023safe}. In \ref*{SI:app:narrow_prior}~\cite{SI}, we assess the robustness of the model-comparison results by fixing the priors of the model-specific parameters and varying the prior widths of the remaining parameters. 

\section{Validating the framework with synthetic data}
\label{sec:bench}

Before applying the framework to experimental data, we first assess its validity using synthetic trajectories generated from the candidate models themselves. This benchmark serves two purposes. First, it tests whether the proposed comparison framework, based on position-increment likelihoods, can recover the true generating model when the data are sampled sufficiently finely. Second, it clarifies how temporal coarse-graining degrades model identifiability, a behavior also observed later in the application to cell-motility data.

Throughout this section, we generate trajectories from known parameter settings, discard the latent velocities, and perform inference using only the sampled position series. We then compare the model evidences of the four candidate models and summarize the results using log Bayes factors, as given in Eq.~\eqref{eq:lbf}. For simplicity, we denote the posterior probability $p(\mathcal{M}_i | \mathcal{D})$ of model $\mathcal{M}_i$ by $p_{\mathcal{M}_i}$; for example, $p_{\rm BM}$ denotes the posterior probability of the BM model. 

\begin{figure}[h]
    \centering
    \includegraphics[width=1.0
    \linewidth]{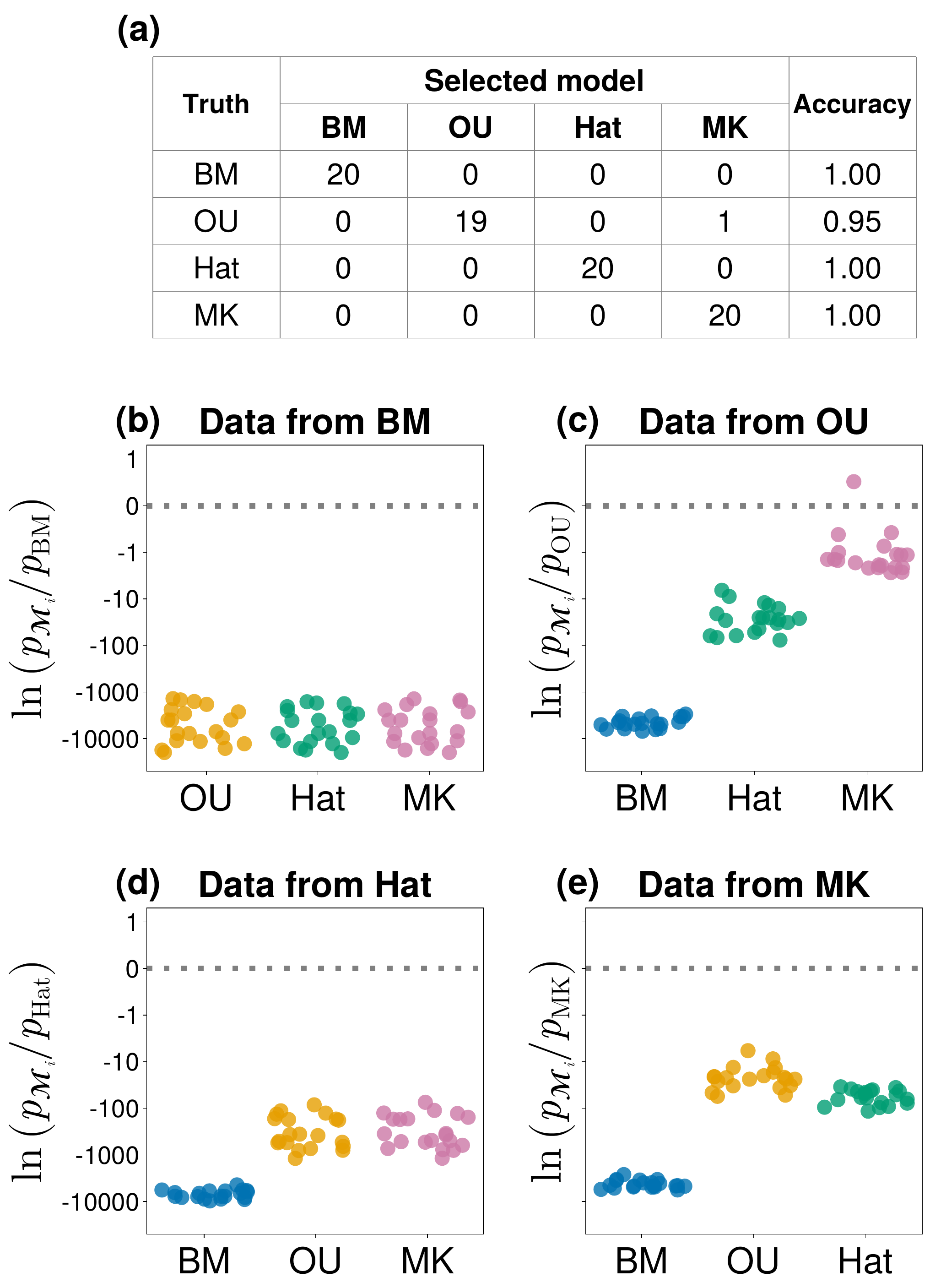}
    \caption{Results of model comparison from synthetic trajectories. We generate $20$ trajectories from each of the four candidate models, with parameters drawn independently for each trajectory. For the BM, OU, and Mexican-hat models, each parameter is drawn from $\mathcal{L}(0.3, 3.0)$. For the Mexican-hat model, we retain only parameter sets satisfying $D/(\gamma v_r^4) < 0.3$; this criterion is discussed in the main text. For the memory kernel model, $D$ and $s$ are drawn from $\mathcal{L}(0.3, 3.0)$, $d_1$ from $\mathcal{L}(0.1, 0.3)$, and $u$ from $\mathcal{L}(10.0, 20.0)$. All trajectories are generated with $T=200$ and $\Delta t=0.1$. Panel (a) shows how often each candidate model is selected for trajectories generated from each model. Panels (b), (c), (d), and (e) show the log Bayes factors for trajectories generated from the BM, OU, Mexican-hat, and memory kernel models, respectively. In each panel, the log Bayes factors are defined relative to the generating model. The gray dotted line marks a log Bayes factor of zero; points below this line indicate that the generating model is favored over the alternative model and is therefore correctly selected.} 
    \label{fig:verify_framework_summary}
\end{figure}

\subsection{Model identification}

We evaluate the validity of our framework by performing model comparison for synthetic trajectories generated from each candidate model. 
Figure~\ref{fig:verify_framework_summary} summarizes the results: panel (a) shows the number of trajectories assigned to each candidate model, whereas panels (b), (c), (d), and (e) show the corresponding log Bayes factors relative to the generating model. 
All trajectories are generated with $T=200$ and $\Delta t=0.1$, using parameter values randomly chosen as described in the figure caption.
For model comparison and parameter inference, we use the priors listed in Table~\ref{tab:prior}. 


Figure~\ref{fig:verify_framework_summary} (a) indicates that the framework reliably identifies the generating model for trajectories generated from the BM, Mexican-hat, and memory kernel models. For trajectories generated from the OU model, the framework correctly selects the OU model in $95\%$ of cases, with the remaining trajectory assigned to the memory-kernel model. This ambiguity is expected because both models define linear Gaussian dynamics, and the memory-kernel model reduces to the OU model in certain limits, for example $A \to 0$ or, for fixed $A$, $a \to \infty$. Importantly, these misclassifications are accompanied by log Bayes factors $\ln (p_{\rm MK}/p_{\rm OU})$ close to zero, as shown in Fig.~\ref{fig:verify_framework_summary} (c), indicating weak evidence rather than a confident preference for the wrong model.

The parameter ranges used for data generation in Fig.~\ref{fig:verify_framework_summary} are described in the figure caption. In particular, for the Mexican-hat model, we retain only parameter sets satisfying $D/(\gamma v_r^4) < 0.3$. This criterion is motivated by comparing the width of the stationary distribution~\eqref{eq:Mhat_stat_dist} around $\|\bm v\|=v_r$ with the preferred speed $v_r$ itself. A Laplace approximation of Eq.~\eqref{eq:Mhat_stat_dist} around $\|\bm v\|=v_r$ gives the local radial variance $D/(2\gamma v_r^2)$. Therefore, resolving the peak at $\|\bm v\|=v_r$ in finite-size data requires $\sqrt{\frac{D}{2\gamma v_r^2}} \ll v_r$, which corresponds to requiring $D/(\gamma v_r^4)$ to be sufficiently small. We therefore impose the upper bound for $D/(\gamma v_r^4)$ when generating Mexican-hat trajectories. 

We summarize the relative errors of the estimated parameters $\hat{\bm{\theta}}_i$ in Fig.~\ref*{fig:wide_prior_paraest} of Ref.~\cite{SI}. The MAP estimates are generally accurate for the BM, OU, and Mexican-hat models, whereas larger errors occur for the memory-kernel model, particularly for the smaller damping rate $d_1$, whose estimation can be affected by the noisier long-lag behavior of the VACF. We also repeat the synthetic-data benchmark with narrower priors and find that the model-comparison accuracy is largely unchanged; see Fig.~\ref*{fig:narrow_prior} of Ref.~\cite{SI}.

\subsection{Sampling-interval dependence of log Bayes factors}

\begin{figure}
    \centering
    \includegraphics[width=0.9\linewidth]{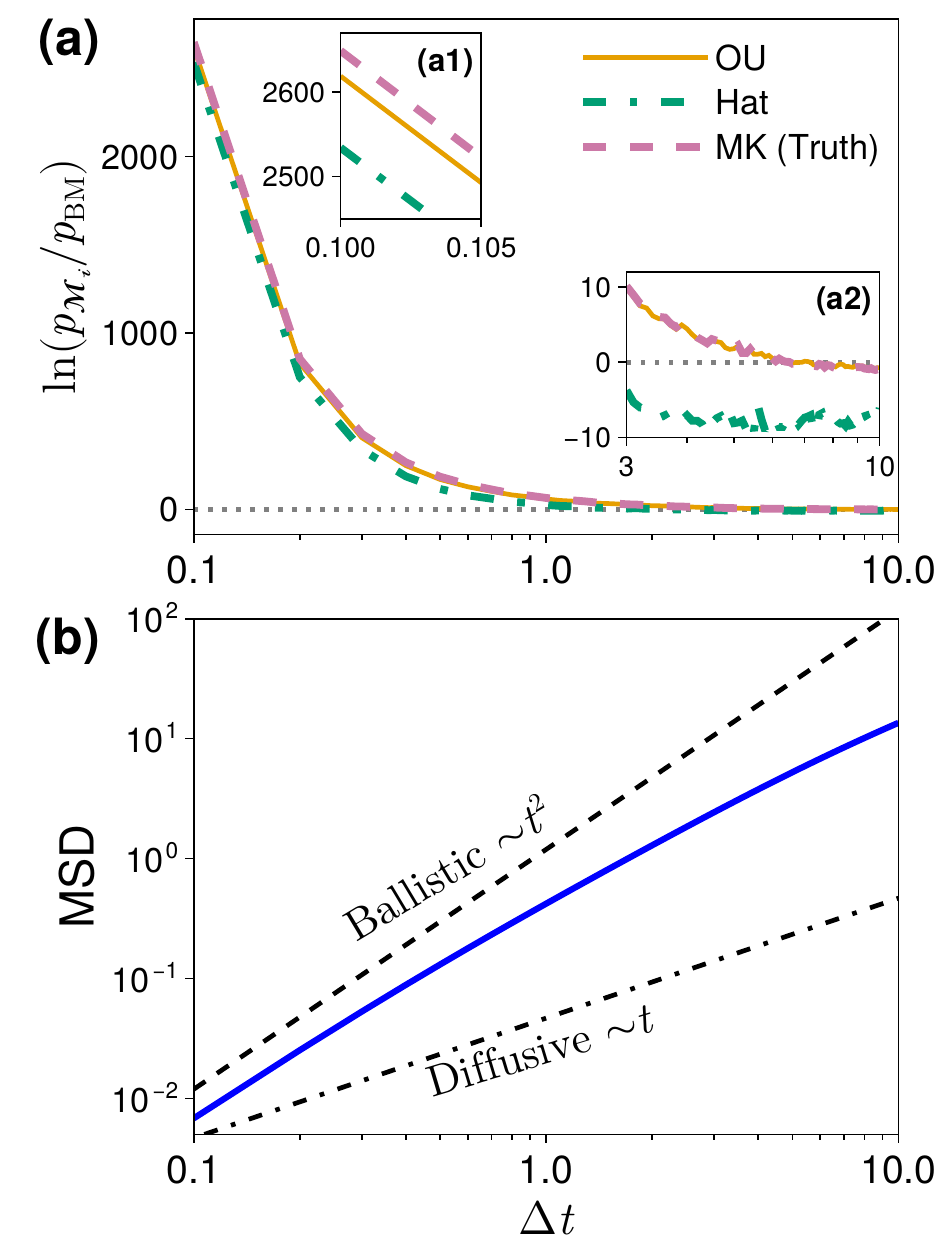}
    \caption{Dependence of model comparison on the sampling interval. (a) We perform model comparison for synthetic trajectories generated from the memory kernel model and plot the log Bayes factors relative to the BM model, $\log(p_{\mathcal{M}_i}/p_{\mathrm{BM}})$, as functions of the sampling interval $\Delta t$.
    The original trajectories are generated with $T=200$ and $\Delta t=0.01$, and trajectories at coarser sampling intervals are obtained by subsampling. We fix the parameters for the memory kernel model at $\gamma=3.5$, $a=1.0$, $A=2.0$, and $D=1.0$, and generate $20$ trajectories with different random seeds. We plot the mean over these $20$ runs. The solid yellow, green dash-dotted, and pink dashed lines represent the log Bayes factors of the OU, Mexican-hat, and memory kernel models, respectively. The inset panels (a1) and (a2) provide enlarged views of the results at small and large sampling intervals, respectively. (b)~MSD for the memory kernel model [Eq.~\eqref{eq:fly_msd}], calculated using the same parameters as those used to generate the data in panel~(a). The dashed and dash-dotted lines indicate reference scalings for ballistic and diffusive motion, respectively.
    }
    \label{fig:comp_artificial_meanstd}
\end{figure}

Next, we examine how temporal resolution affects model identifiability. Figure~\ref{fig:comp_artificial_meanstd} (a) shows the log Bayes factor relative to the BM model as functions of the sampling interval $\Delta t$ for trajectories generated from the memory kernel model. To relate the model-comparison results to the observable trajectory statistics, Fig.~\ref{fig:comp_artificial_meanstd} (b) shows the corresponding MSD [i.e., Eq.~\eqref{eq:fly_msd}] evaluated using the same parameter values. Additional results for trajectories generated from the OU and Mexican-hat models are presented in Fig.~\ref*{fig:comp_artificial_meanstd_other} of Ref.~\cite{SI} and show qualitatively similar trends.

At fine sampling intervals, the model evidence generally favors the true generating model, as seen in Fig.~\ref{fig:comp_artificial_meanstd} (a). This result indicates that, when trajectories are sufficiently well resolved, the proposed framework can extract latent dynamical information from position-only observations. The difference of log Bayes factors between the OU and memory kernel models is nevertheless relatively small, which is consistent with the close relationship between these two linear Gaussian models.

As $\Delta t$ increases, however, the log Bayes factors relative the BM model approach zero, and the model ranking becomes less decisive. In particular, Fig.~\ref{fig:comp_artificial_meanstd}  (a2) shows that even the true memory kernel model, as well as its similar OU model, become increasingly difficult to distinguish from BM at coarse sampling intervals. 
Figure~\ref{fig:comp_artificial_meanstd} (b) helps explain this trend: as the time lag $\Delta t$ increases, the MSD departs from the ballistic reference (the dashed line) and approaches the diffusive reference (the dash-dotted line). Consequently, coarse sampling emphasizes the diffusive regime, making the memory-kernel dynamics more difficult to distinguish from BM.
This scale-dependent loss of identifiability is an important consideration for the experimental analysis of cell motility in Sec.~\ref{sec:experiment}, where the sampling interval limits the dynamical complexity that can be statistically supported by the data.

We also use this numerical experiment as a consistency check of the likelihood approximation scheme [Eqs.~\eqref{eq:likeli_nonlin_secant}, \eqref{eq:likeli_nonlin_approx}], which we employ for the Mexican-hat model. Sepecifically, we derive the approximate likelihood for the OU model, where the exact likelihood is available, and compute the log Bayes factors over the same range of $\Delta t$ using both the approximate and exact likelihoods. As shown in Fig.~\ref*{fig:OUvsOUexact} of Ref.~\cite{SI}, the two results agree well over a wide range of $\Delta t$. This agreement supports the application of the approximate likelihood to nonlinear models, for which an exact position-only likelihood is generally unavailable.

\section{Application to motility data}
\label{sec:experiment}
After benchmarking our framework, we apply it to experimentally obtained trajectories of {\it D.~discoideum} cells. We first briefly summarize the experimental setup and pre-processing for single-cell tracking data and then perform a preliminary analysis using standard statistics, such as the MSD, to characterize the data and determine the subsampling intervals for subsequent model selection. Finally, we present model-comparison results at three sampling intervals $\Delta t$, discuss how model identifiability changes under temporal coarse-graining, and evaluate the reliability of the selected models using the log Bayes factors and predicted trajectory statistics.

\subsection{Experimental setup and pre-processing}
We use single-cell tracking trajectories of the DdB wild-type strain of {\it D.~discoideum}. Cell positions are recorded every $5\mathrm{sec}$, and the longest trajectories contain $1088$ data points, corresponding to nearly \qty{1.5}{\hour}.

For the analysis and model comparison below, we retain only trajectories with at least 257 recorded positions, corresponding to observation times longer than $\qty{1280}{\second}$. This cutoff ensures that, even after temporal subsampling to the coarsest interval used for the subsequent model comparison (\qty{640}{\second}), each trajectory contains at least three data points. After filtering, the DdB dataset contains 261 trajectories. Additional details of the experimental setup and data pre-processing are provided in Appendix~\ref{app:exp_detail}. 

The DdB stain is closely related to the original {\it D.~discoideum} isolate NC4~\cite{bloomfield2008widespread}.  Corresponding model-comparison results for the AX2 wild-type strain, with and without bacteria, and for the NF1 knockout strain with bacteria are presented in Supplementary Information.

\subsection{Preliminary analysis}
Before carrying out Bayesian model comparison, we examine these standard trajectory statistics and finite-difference velocity distributions as a preliminary characterization of the DdB data. The purpose of this step is not to select models directly from summary statistics, but to identify representative sampling intervals at which different effective descriptions are expected to emerge and to provide intuition for the subsequent evidence-based analysis.

\begin{figure}
    \centering    
    \includegraphics[width=1.\linewidth]{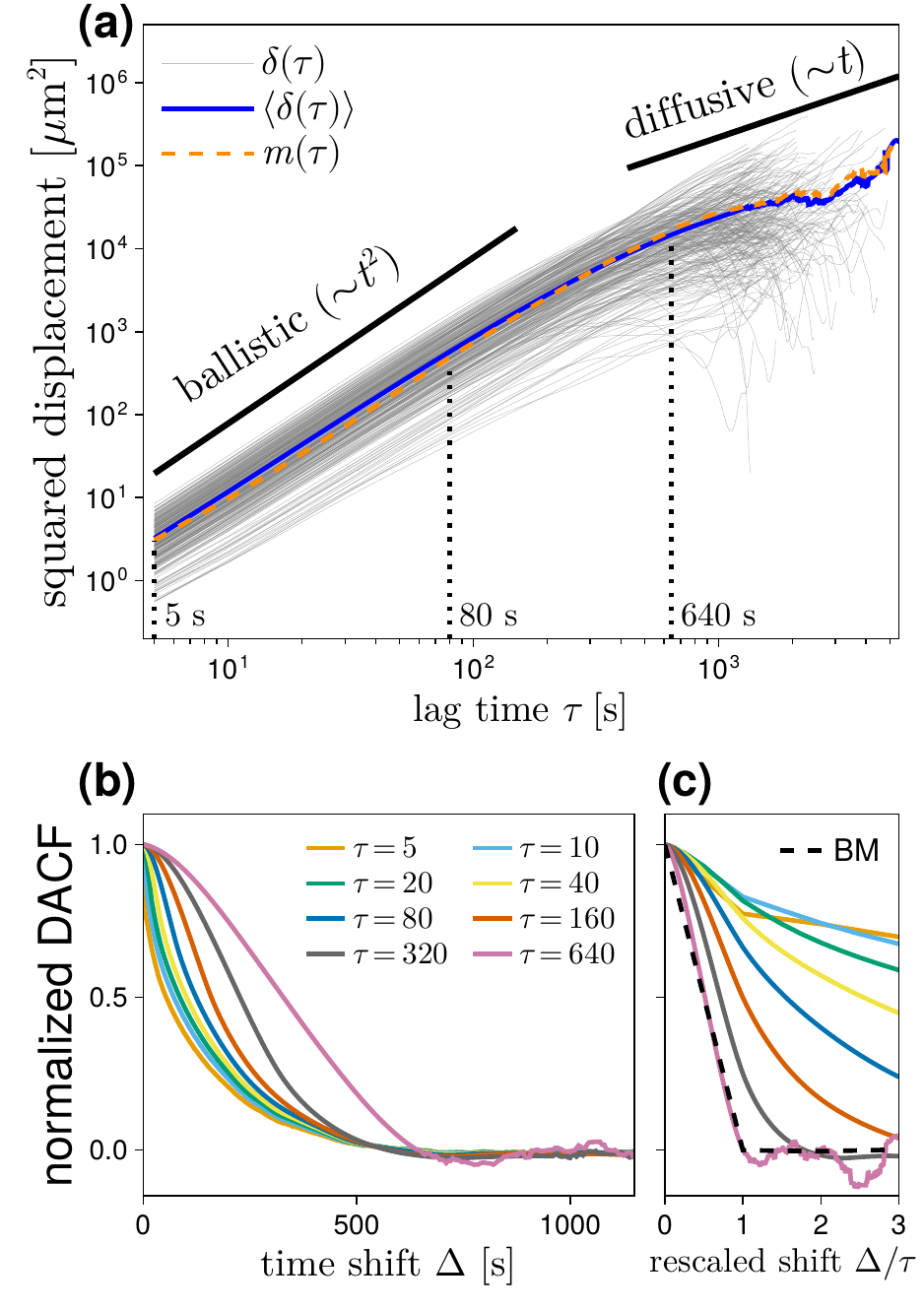}
    \caption{Preliminary analysis of DdB cell trajectories. (a)~Ensemble-averaged MSD [$m(\tau)$; orange dashed line], together with the TAMSD [$\delta(\tau)$; gray thin lines] for individual trajectories and their ensemble average [$\langle \delta(\tau) \rangle$; blue line]. The two black solid lines indicate reference scalings for ballistic and diffusive motion. The black dotted lines show the three time scales (i.e., $\tau = 5,\ 80$, and $\qty{640}{\second}$) used for the subsequent model comparison. 
    (b)~Ensemble averaged normalized DACF $\langle \tilde S(\Delta,\tau) \rangle$ for several fixed lag times $\tau$. (c)~The same curves plotted against the rescaled time shift $\Delta/\tau$. The black dashed line shows the normalized DACF computed from synthetic trajectories generated by the BM model. }
    \label{fig:DdB_data}
\end{figure}

For a discrete trajectory $\{\bm r_i\}_{i=0}^{M}$ sampled at interval $\Delta t$, we evaluate displacement statistics at lag time $\tau = \ell \Delta t$, where $\ell\ge1$ is an integer. In addition to the ensemble-averaged MSD 
\begin{equation}
    m(\tau) \coloneqq \langle \|\bm r_{\ell}-\bm r_0 \|^2 \rangle ,
\end{equation}
where the brackets denote an average over trajectories, 
we compute the time-averaged MSD (TAMSD) for each trajectory as
\begin{equation}
    \delta(\tau) \coloneqq \frac{1}{M-\ell+1}\sum_{i=0}^{M-\ell}\|\bm r_{i+\ell}-\bm r_i\|^2 \, ,  
\end{equation}
and its ensemble average $\langle \delta(\tau) \rangle$. Figure~\ref{fig:DdB_data} (a) shows $m(\tau)$, the individual TAMSDs $\delta(\tau)$, and their ensemble average $\langle \delta(\tau) \rangle$ for the DdB trajectories. These MSD statistics indicate ballistic scaling at short timescales ($\tau \lesssim \qty{500}{\second}$) and diffusive scaling at longer timescales ($\tau \gtrsim \qty{500}{\second}$).

We further evaluate the displacement autocorrelation function (DACF)~\cite{grossmann2024non, albrecht2026likelihood}. For a time shift $\Delta = k \Delta t$, with integer $0\le k\le M-\ell$, the DACF is defined as
\begin{align}
    S(\Delta, \tau)
    \coloneqq 
    \frac{\displaystyle\sum_{i=0}^{M-\ell-k}
    (\bm r_{i+\ell}-\bm r_i)\cdot(\bm r_{i+\ell+k}-\bm r_{i+k})}{M-\ell-k+1} \, ,
\end{align}
where the dot denotes the inner product of two vectors. The DACF measures the correlation between two displacements of duration $\tau$ separated by the time shift $\Delta$. For each fixed $\tau$, we examine how $S(\Delta, \tau)$ depends on $\Delta$. In particular, we focus on the $\Delta$ dependency of the normalized DACF 
\begin{equation} 
    \tilde S(\Delta, \tau) \coloneqq S(\Delta, \tau) / S(0, \tau),
\end{equation}
and its ensemble average $\langle \tilde S(\Delta, \tau) \rangle$. Figures~\ref{fig:DdB_data}~(b) and~\ref{fig:DdB_data}~(c) show $\langle \tilde S(\Delta,\tau) \rangle$ as a function of the time shift~$\Delta$ and the rescaled time shift~$\Delta/\tau$, respectively. As shown in panel~(c), the normalized DACF approaches the Brownian-motion reference curve (black dashed line) as the lag time $\tau$ increases. At $\tau = \qty{640}{\second}$, the normalized DACF is nearly identical to the BM reference, indicating that the motion is close to effective normal diffusion at this timescale.

\begin{figure}
    \centering
    \includegraphics[width=.9\linewidth]{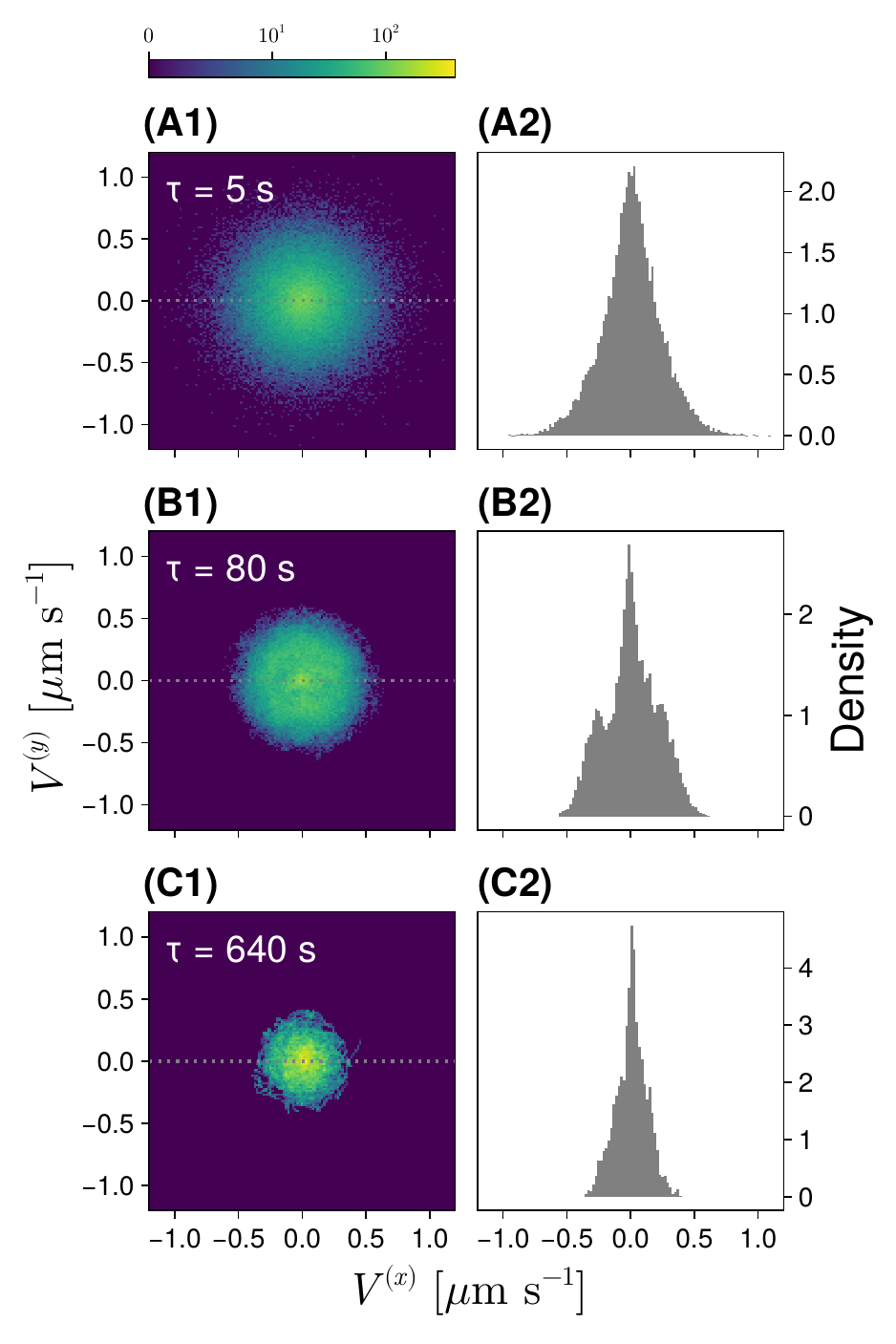}
    \caption{Velocity distributions of DdB cells for different finite-difference intervals $\tau = \ell \Delta t$. For each value of $\tau$, velocities are computed from the position trajectories as $\bm{V}_i=(\bm{r}_{i+\ell}-\bm{r}_i)/\tau$. The left column shows two-dimensional histograms of the sampled velocities in the $(V^{(x)},V^{(y)})$ plane. The right column shows one-dimensional histograms of velocity samples near $V^{(y)}=0$; more precisely, samples satisfying $|V^{(y)}| < \qty{0.05}{\micro\meter\per\second}$ are included. These plots illustrate how the apparent velocity distribution changes with the interval $\tau$. Panels~(A1) and~(A2) correspond to $\tau = \qty{5}{\second}$, panels~(B1) and~(B2) to $\tau = \qty{80}{\second}$, and panels~(C1) and~(C2) to $\tau = \qty{640}{\second}$.}
    \label{fig:DdB_veldis}
\end{figure}

Finally, we examine the velocity distributions for different intervals $\tau = \ell \Delta t$. In Fig.~\ref{fig:DdB_veldis}, we plot the distributions of the finite-difference velocity $\bm V_i=(\bm r_{i+\ell}-\bm r_i)/\tau$ computed at three intervals, $\tau = 5,\ 80$, and $\qty{640}{\second}$. 
We use the same notation $\bm V_i$ as in Eq.~\eqref{eq:secant-velocity}, but for $\ell>1$ it denotes a coarse-grained finite-difference velocity rather than the original one-step secant velocity. Although the velocity distributions at the short and long intervals are unimodal [panels~(a) and~(c), respectively], the distribution at the intermediate interval retains a multimodal structure [panel~(b)] that is qualitatively compatible with preferred-speed dynamics. This observation is consistent with a previous report on another strain (AX4) of {\it D.~discoideum}~\cite{li2008persistent} and motivates closer examination of the dynamics around $\tau=\qty{80}{\second}$.

Based on the above analyses, we use the original sampling interval, $\Delta t = \qty{5}{\second}$, together with two additional coarse-grained intervals: $\Delta t = \qty{80}{\second}$, where a subset of cells exhibits preferred-speed-like behavior [Fig.~\ref{fig:DdB_veldis} (b)], and $\Delta t = \qty{640}{\second}$, where the trajectories are close to the BM reference [Fig.~\ref{fig:DdB_data} (c)]. To construct trajectories at the two coarser sampling intervals, we subsample the original trajectories every $16$ or $128$ data points, at the expense of reducing the number of data points. We then perform the subsequent model comparison and data analysis using trajectories at these three sampling intervals. 


\subsection{Results of model comparison}
\label{sec:dicty_comparison}

\begin{figure}
    \centering
    \includegraphics[width=1.\linewidth]{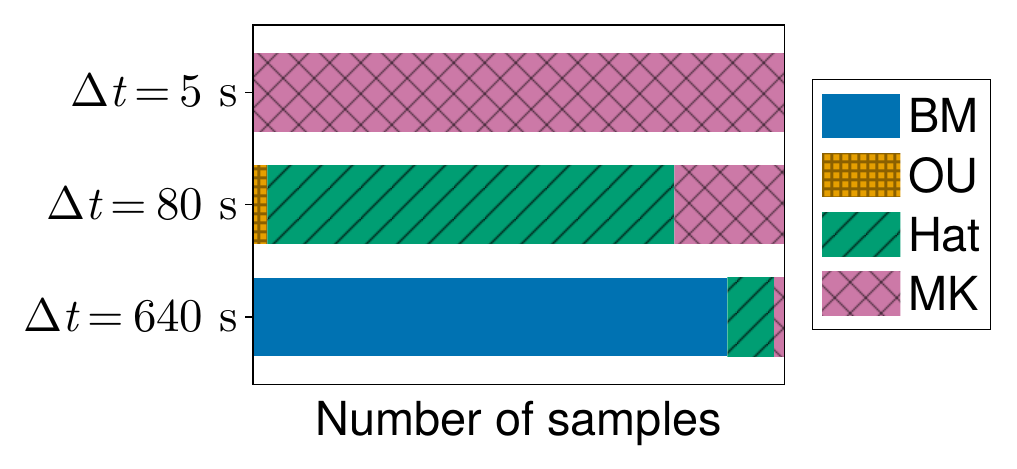}
    \caption{Counts of selected models across sampling intervals.
    For each sampling interval $\Delta t$ [\qty{}{\second}], trajectories are classified according to the model with the largest marginal likelihood~\eqref{eq:ml} among the BM, OU, Mexican-hat (Hat), and memory kernel (MK) models. Stacked bars indicate the number of trajectories assigned to each model for the DdB cell dataset ($261$ trajectories in total). 
    }
    \label{fig:DdB_comp_summary}
\end{figure}

We apply the proposed comparison framework to cell motility trajectories of the {\it D.~discoideum} DdB strain.
Figure~\ref{fig:DdB_comp_summary} summarizes the distribution of selected models across the sampling intervals $\Delta t$.

At the shortest sampling interval ($\Delta t = \qty{5}{\second}$), all trajectories are classified as the memory-kernel model, suggesting that the finest-resolution data contain short-time structure not captured by models without an explicit memory kernel. 
This interpretation should be treated cautiously, however, because unmodeled observational noise may also affect short-time increments.
At the intermediate sampling interval ($\Delta t = \qty{80}{\second}$), 
all three second-order models are selected,
reflecting heterogeneous dynamical behavior across trajectories.
At the largest sampling interval ($\Delta t = \qty{640}{\second}$), most trajectories are classified as the BM model, suggesting that more complex dynamics become indistinguishable from simpler effective descriptions under temporal coarse-graining.

\begin{figure}
    \centering
    \includegraphics[width=1.0\linewidth]{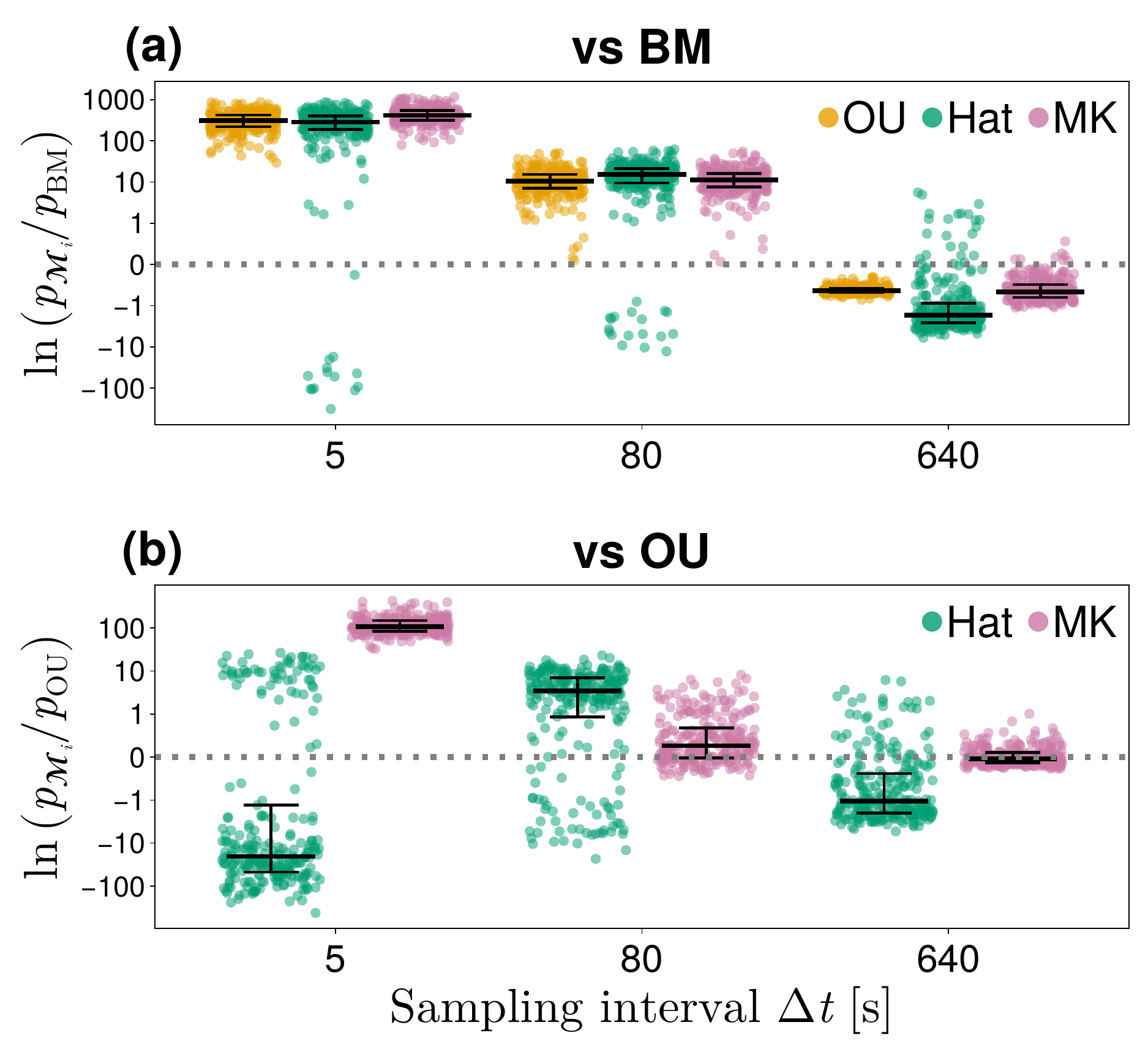}
    \caption{Log Bayes factors for model comparison from DdB trajectories at different sampling intervals. The left, middle, and right groups on the horizontal axis correspond to $\Delta t=5,\ 80$, and $\qty{640}{\second}$, respectively. (a) Log Bayes factors relative to the BM model. The orange, green, and magenda points show $\log(p_{\mathrm{OU}}/p_{\mathrm{BM}})$, $\log(p_{\mathrm{Hat}}/p_{\mathrm{BM}})$, and $\log(p_{\mathrm{MK}}/p_{\mathrm{BM}})$, respectively. (b) Log Bayes factors relative to the OU model. The green and magenda points show $\log(p_{\mathrm{Hat}}/p_{\mathrm{OU}})$ and $\log(p_{\mathrm{MK}}/p_{\mathrm{OU}})$, respectively. Each point corresponds to one trajectory. Black horizontal bars indicate medians, and vertical black bars with caps indicate interquartile ranges.
    }
    \label{fig:DdB_comp_vsBM_vsOU}
\end{figure}

To quantify the statistical support for the model assignments, we examine the distributions of log Bayes factors computed for individual DdB trajectories. Figure~\ref{fig:DdB_comp_vsBM_vsOU}~(a) shows the log Bayes factors of the OU, Mexican-hat, and memory kernel models relative to the BM model. The three models are strongly favored over the BM model at fine temporal resolution $\Delta t = \qty{5}{\second}$ [left group of Fig.~\ref{fig:DdB_comp_vsBM_vsOU}~(a)]. However, the differences in model evidence shrink substantially as $\Delta t$ increases [middle and right groups of Fig.~\ref{fig:DdB_comp_vsBM_vsOU}~(a)], indicating that the model selection becomes less decisive at coarser temporal resolution.

Figures~\ref{fig:DdB_comp_vsBM_vsOU}~(b) show the log Bayes factors of the Mexican-hat and memory kernel models relative to the OU model. At the smallest interval [left group of Fig.~\ref{fig:DdB_comp_vsBM_vsOU}~(b)], the memory kernel model is clearly favored over the OU model.
In contrast, at the intermediate interval [middle group of Fig.~\ref{fig:DdB_comp_vsBM_vsOU}~(b)], the differences in model evidence between the OU and memory kernel models are small, while the majority of trajectories show stronger support for the Mexican-hat model. At the largest interval [right group of Fig.~\ref{fig:DdB_comp_vsBM_vsOU}~(b)], the log Bayes factors lie close to $0$, again indicating that the models become statistically indistinguishable at coarser temporal resolution.

\subsection{Prediction from selected models}
\label{sec:comp_prediction}
\begin{figure}
    \centering
    \includegraphics[width=1.\linewidth]{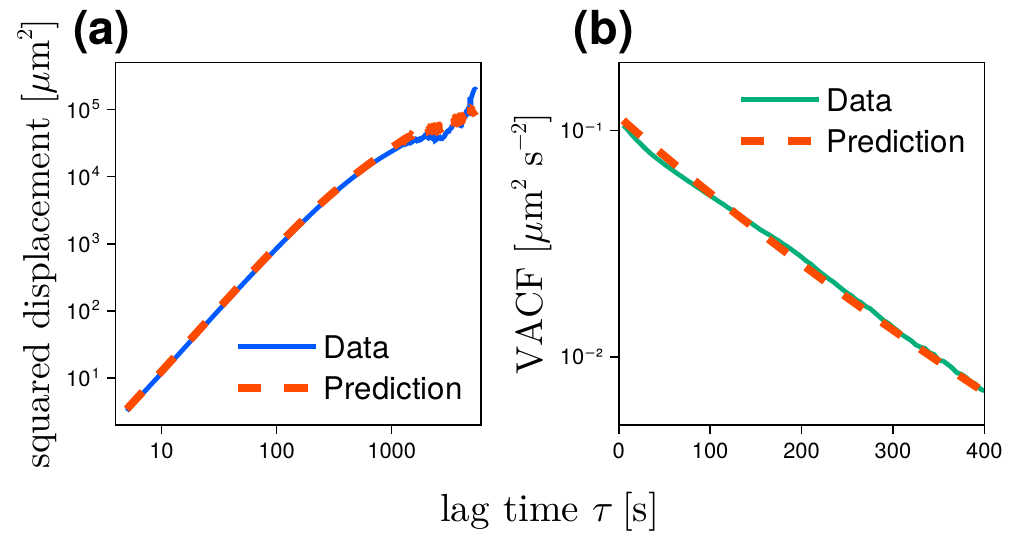}
    \caption{Comparison of empirical trajectory statistics with theoretical predictions obtained from the inferred model parameters. Results are shown for the $261$ DdB trajectories sampled at $\Delta t = \qty{5}{\second}$, for all of which the memory kernel model is selected. In panel (a), the blue solid line shows the ensemble-averaged TAMSD $\langle \delta(\tau) \rangle$ computed from the DdB trajectories, whereas the red dashed line shows the ensemble-averaged MSD prediction, $\langle \hat{m}(\tau) \rangle$, computed using Eq.~\eqref{eq:fly_msd} with the inferred parameters shown in Fig.~\ref{fig:DdB_comp_5sec}. In panel (b), the green solid line shows the ensemble-averaged VACF computed from the data. Because the instantaneous velocity is not directly observed, we calculate the autocorrelation of the secant velocity $\bm V_i$ for each trajectory as $C_V(\tau) \coloneqq (M-\ell)^{-1}\sum_{i=0}^{M-\ell-1}\bm V_{i+\ell}\cdot\bm V_i$ and plot their ensemble average $\langle C_V(\tau) \rangle$. The red dashed line shows the ensemble-averaged VACF prediction, $\langle C_v(\tau) \rangle$, computed using Eq.~\eqref{eq:fly_VACF} with the inferred parameters shown in Fig.~\ref{fig:DdB_comp_5sec}. }
    \label{fig:DdB_comp_5sec_prediction}
\end{figure}

To further assess whether the selected models provide a quantitatively consistent description of the DdB trajectories, we test whether they can reproduce the corresponding trajectory statistics using the inferred parameters. Figure~\ref{fig:DdB_comp_5sec_prediction} shows the results for the finest sampling interval, $\Delta t = \qty{5}{\second}$, where we compare the MSD and VACF computed from the original DdB trajectories with model predictions generated from the selected memory kernel models at their inferred MAP parameter values. The MAP parameter values used for these predictions are shown in Fig.~\ref{fig:DdB_comp_5sec} in Appendix~\ref{app:exp_paraest}. 

The resulting model predictions are shown as red dashed curves in Fig.~\ref{fig:DdB_comp_5sec_prediction} (a) and (b), together with the empirical statistics computed directly from the DdB trajectories (solid curves). The predicted curves closely reproduce the data-based MSD and VACF over the entire range shown. This agreement for two distinct trajectory statistics provides a stringent validation of the selected memory kernel model beyond the Bayes-factor comparison alone. Combined with the large positive log Bayes factors relative to the other models [left groups of Figs.~\ref{fig:DdB_comp_vsBM_vsOU}~(a) and(b)], the successful prediction of both the MSD and VACF strongly supports the memory kernel model at $\Delta t = \qty{5}{\second}$ and indicates the presence of two characteristic time scales at this temporal resolution.

\begin{figure}
    \centering
    \includegraphics[width=1.\linewidth]{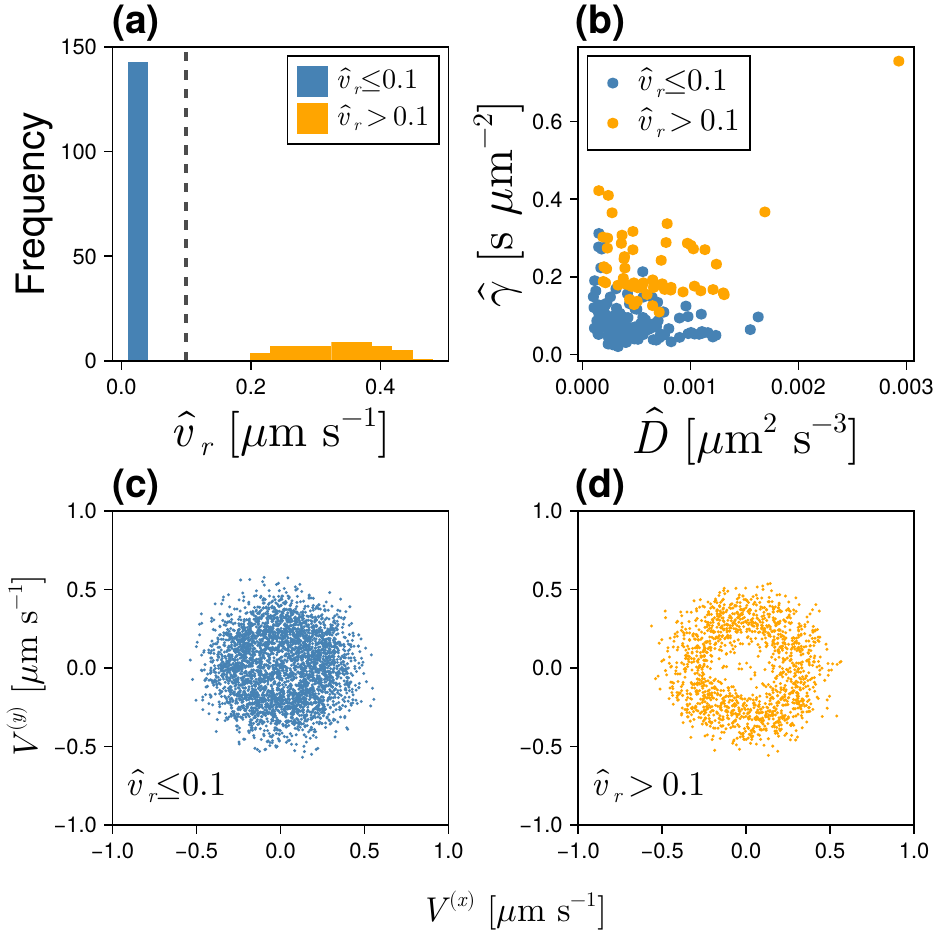}
    \caption{Parameter estimates and velocity distributions for the $200$ DdB trajectories selected as the Mexican-hat model at sampling interval $\Delta t = \qty{80}{\second}$. Panel (a) shows the distribution of the inferred preferred speed $\hat{v}_r$. Throughout this figure, colors indicate whether each trajectory has estimated preferred speed $\hat{v}_r > \qty{0.1}{\micro\meter\per\second}$ (orange) or $\hat{v}_r \leq \qty{0.1}{\micro\meter\per\second}$ (blue). The gray dashed line in panel (a) indicate this separation line $\hat{v_r}=\qty{0.1}{\micro\meter\per\second}$. Panel (b) shows a scatter plot of the inferred parameters $\hat{\gamma}$ and $\hat{D}$. Panels (c) and (d) show the distributions of secant velocities $\bm V$ for trajectories with $\hat{v}_r \leq \qty{0.1}{\micro\meter\per\second}$ and $\hat{v}_r > \qty{0.1}{\micro\meter\per\second}$, respectively. Here, $\bm V$ is computed from adjacent positional increments of the subsampled trajectories at $\Delta t = \qty{80}{\second}$. 
    }
    \label{fig:DdB_comp_80sec}
\end{figure}

At the intermediate sampling interval $\Delta t = \qty{80}{\second}$, most trajectories are selected as the Mexican-hat model (Fig.~\ref{fig:DdB_comp_summary}). 
We therefore examine the inferred parameters and secant-velocity distributions for the trajectories selected as the Mexican-hat model at this sampling interval (Fig.~\ref{fig:DdB_comp_80sec}). The trajectories are divided into two groups according to the inferred preferred speed $\hat{v}_r$, with blue indicating $\hat{v}_r \leq \qty{0.1}{\micro\meter\per\second}$ and orange indicating $\hat{v}_r > \qty{0.1}{\micro\meter\per\second}$, as shown in panel (a). Panel (b) shows that the MAP estimates of the remaining parameters, $\hat{D}$ and $\hat{\gamma}$, occupy different ranges in these two groups. Panels (c) and (d) show the corresponding distributions of the secant velocity. In particular, the subgroup with $\hat{v}_r > \qty{0.1}{\micro\meter\per\second}$ displays a ring-shaped velocity distribution [panel (d)], consistent with a finite preferred speed and with the characteristic structure of the Mexican-hat model. This agreement between the inferred parameter values and the observed velocity-distribution structure provides an additional consistency check of the model-selection result.


\begin{figure}
    \centering
    \includegraphics[width=1.\linewidth]{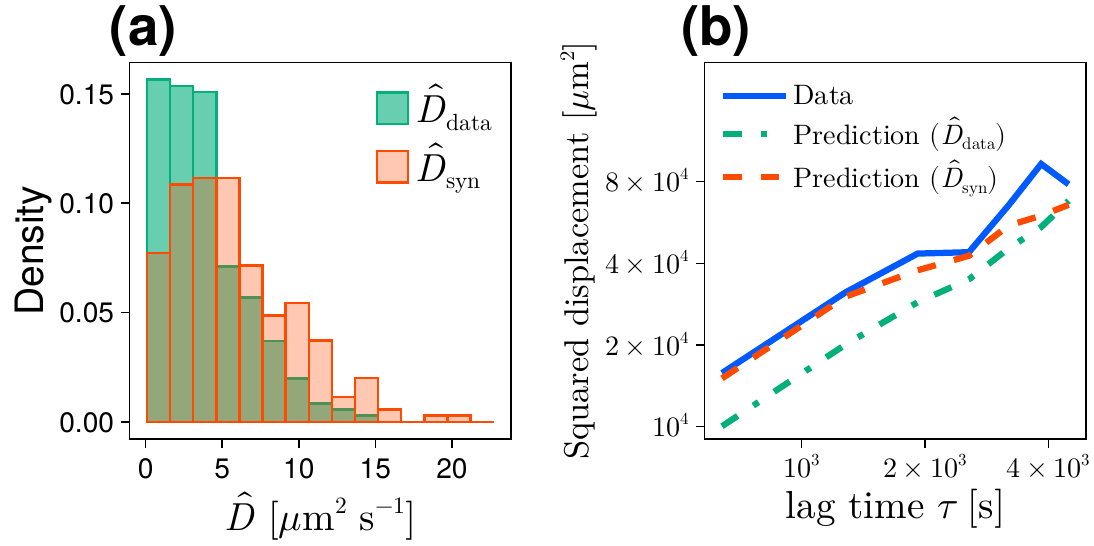}
    \caption{Distribution of inferred parameters and MSDs for the $233$ trajectories selected as BM at the largest sampling interval, $\Delta t = \qty{640}{\second}$. Panel~(a) shows the distributions of the MAP estimates of the diffusion coefficient. The green histogram represents the estimates $\hat{D}_{\mathrm{data}}$ obtained directly from the subsampled DdB trajectories. The red histogram represents the estimates $\hat{D}_{\mathrm{syn}}$ obtained from long synthetic trajectories generated at $\Delta t = \qty{640}{\second}$ using the memory kernel models and MAP parameters inferred from the corresponding trajectories at $\Delta t = \qty{5}{\second}$. Panel~(b) compares the ensemble-averaged TAMSD, $\langle \delta(\tau) \rangle$, of the subsampled DdB trajectories (blue solid line) with the BM-based MSD predictions $\langle \hat{m}(\tau) \rangle$ computed using Eq.~\eqref{eq:BM_msd}. The green dash-dotted line is calculated using $\hat{D}_{\mathrm{data}}$, whereas the red dashed line is calculated using $\hat{D}_{\mathrm{syn}}$. 
    }
    \label{fig:DdB_comp_640sec}
\end{figure}

At the largest sampling interval, $\Delta t = \qty{640}{\second}$, the second-order models are no longer clearly distinguishable from the BM model for most trajectories [right group of Fig.~\ref{fig:DdB_comp_vsBM_vsOU} (a)]. We therefore focus on the trajectories classified as Brownian motion at this sampling interval. 
The diffusion coefficients inferred directly from the coarsely sampled DdB trajectories, $\hat D_\mathrm{data}$, are shown in green in Fig.~\ref{fig:DdB_comp_640sec}~(a). The corresponding MSD prediction, shown by the green dash-dotted curve in panel (b), exhibits a modest discrepancy from the empirical MSD shown by the blue solid curve. This discrepancy is considered to arise mainly from uncertainty in estimating the diffusion coefficient, owing to the limited number of data points after subsampling. To obtain more stable estimates, we generate long trajectories at $\Delta t = \qty{640}{\second}$ from the memory kernel models inferred at $\Delta t = \qty{5}{\second}$ and estimate the effective BM diffusion coefficients from these trajectories. The resulting estimates, $\hat D_\mathrm{syn}$, are shown in red in panel (a), and the corresponding MSD prediction is shown by the red dashed curve in panel (b). Its improved agreement with the empirical MSD supports an effective normal-diffusion description at this coarse-grained time scale and suggests that the discrepancy in the direct prediction is attributable to the uncertainty in parameter estimation.

Overall, the comparison results for the DdB trajectories can be summarized in two points. First, as shown in Fig.~\ref{fig:DdB_comp_summary}, model selection depends strongly on the temporal resolution of the observed trajectories. Temporal coarse-graining makes the model assignments increasingly ambiguous, as reflected in the log Bayes factors approaching to $0$ (Fig.~\ref{fig:DdB_comp_vsBM_vsOU}), and the trajectories become effectively closer to normal diffusion. This sampling-dependent loss of identifiability is not unique to the DdB cell trajectories. Applying the model comparison to other cell strains yields the same qualitative trend, as shown in Figs.~\ref*{fig:other_comp_summary} and~\ref*{fig:other_comp_vsBM_vsOU} of Ref.~\cite{SI}. Second, at the shortest and intermediate sampling intervals, the favored candidate models are the memory kernel model for $\Delta t = \qty{5}{\second}$ and the Mexican-hat model for $\Delta t = \qty{80}{\second}$. These assignments are supported both by the log Bayes factors (Fig.~\ref{fig:DdB_comp_vsBM_vsOU}) and by comparisons between empirical statistics and model-based quantities [Fig.~\ref{fig:DdB_comp_5sec_prediction} and Figs.~\ref{fig:DdB_comp_80sec} (c) and (d)]. Nonetheless, the selection of the memory-kernel model at $\Delta t = \qty{5}{\second}$ should be interpreted with caution, because unmodeled short-time fluctuations, such as observational noise, can introduce additional short time scales into the observed positional trajectories. We therefore interpret this result as evidence for the presence of multiple time scales in the trajectories, rather than as proof that the memory-kernel model uniquely describes DdB cell motion.



\section{Discussion}
\label{sec:dis}
We have presented a Bayesian framework for comparing second-order Langevin models from position-only trajectories. The central methodological point is to base model comparison on marginal likelihoods of the increment process obtained after integrating out latent velocities, thereby avoiding the bias associated with naive velocity reconstruction. Within this framework, exact increment likelihoods can be used for linear Gaussian models, while previously proposed approximate likelihoods~\cite{albrecht2024inferring} can be incorporated for the nonlinear Mexican-hat model. 

The synthetic-data benchmarks in Sec.~\ref{sec:bench} that the proposed framework can reliably recover the generating model when the sampling interval is sufficiently short. At the same time, they also clarify a key limitation: as the temporal resolution becomes coarse, distinct dynamical models can become statistically indistinguishable at the level of position-only observations. In this sense, the framework is useful not only for selecting among competing models, but also for quantifying the practical loss of identifiability induced by temporal coarse-graining.

For the {\it D.~discoideum} trajectories, the supported model class depends strongly on the sampling interval. As shown in Figs.~\ref{fig:DdB_comp_summary}, the memory kernel model is supported for a substantial fraction of trajectories at the shortest sampling interval ($\Delta t = \qty{5}{\second}$). Importantly, the models inferred at this interval also reproduce both the MSD and VACF computed from the experimental trajectories (Fig.~\ref{fig:DdB_comp_5sec_prediction}). This agreement provides an additional consistency check of the model-selection result beyond the Bayes factors alone. At the intermediate interval ($\Delta t = \qty{80}{\second}$), simpler models are also selected, whereas at the largest interval ($\Delta t = \qty{640}{\second}$) the dynamics are no longer distinguishable from normal diffusion, represented here by the BM model. We interpret these results primarily as evidence for sampling-dependent effective descriptions supported both by model comparison and by the successful prediction of trajectory statistics, rather than as proof that a single model uniquely characterizes all cell trajectories across all temporal scales.

The frequent selection of the Mexican-hat model at $\Delta t = \qty{80}{\second}$ is particularly suggestive. A linear Gaussian memory-kernel model, even if supported at $\Delta t = \qty{5}{\second}$, cannot by temporal coarse-graining alone generate the preferred-speed dynamics captured by the Mexican-hat model. This observation points to the need for models that can combine multiple time scales at fine temporal resolution, as proposed previously~\cite{selmeczi2005cell,takagi2008functional}, with nonlinear velocity statistics such as the ring-shaped velocity distribution observed after temporal coarse-graining [Fig.~\ref{fig:DdB_comp_80sec}~(d)]. 

As discussed in Sec.~\ref{sec:comp_prediction}, model selection at the finest sampling interval should be interpreted with caution. The simultaneous agreement of the predicted MSD and VACF with the corresponding experimental statistics (Fig.~\ref{fig:DdB_comp_5sec_prediction}) supports the presence of multiple characteristic time scales in the finely sampled trajectories and shows that the inferred memory kernel models capture important dynamical features of the data. However, this agreement does not establish that the memory kernel model is the uniquely correct mechanistic description of {\it D.~discoideum} motility. In particular, additional sources of fluctuation, including observational noise, may affect model preference at short time scales. Incorporating observational noise into the present framework is therefore an important direction for future research, although the corresponding likelihood calculation is expected to be challenging, especially for nonlinear models. 

To assess the reliability of the importance-sampling estimates, we monitor the effective sampling size (ESS) defined in Eq.~\eqref{eq:ess}. The ESS values for the synthetic and experimental trajectories are summarized in \ref*{app:ess} of Ref.\cite{SI}. Although the ESS is typically $\gtrsim 100$, it can decrease to $\sim 10$ for some trajectories, particularly in model-mismatch cases involving the Mexican-hat and memory kernel models. Because low ESS values indicate less reliable importance-sampling estimates~\cite{martino2017effective,elvira2022rethinking}, we interpret model rankings cautiously when the corresponding log Bayes factors are close to zero, where numerical errors may affect the ordering. We therefore assess model support using both the selected model and the magnitude of its log Bayes factors. More adaptive methods, such as exchange Monte Carlo~\cite{hukushima1996exchange,nagata2012bayesian,kato2025bayesian,kashiwamura2026uncertainty}, may further improve marginal-likelihood estimation, albeit at greater computational cost.

In summary, this study presents two main contributions. First, we formulate Bayesian model comparison for second-order stochastic dynamics directly from position-only observations, using the non-Markovian increment process rather than reconstructed velocities. This formulation enables evidence-based comparison across multiple model classes under partial observation by combining exact increment likelihoods for linear Gaussian models with an approximate likelihood for nonlinear dynamics. Although motivated by cell motility, the approach is applicable more broadly to stochastic dynamics in general, especially when only part of the dynamical state is observed.

Second, applying the framework to experimental {\it D.~discoideum} trajectories shows that model preference depends strongly on the sampling interval. As the trajectories are temporally coarse-grained, the favored description changes from a linear Gaussian model with multiple time scales, to nonlinear preferred-speed dynamics, and finally to normal diffusion. The inferred models also reproduce the corresponding trajectory statistics, providing predictive support for these scale-dependent descriptions beyond the marginal-likelihood comparison alone. This scale-dependent transition indicates that different dynamical features become visible at different temporal resolutions, and highlights the need for model comparison methods that explicitly account for observation scale.





\begin{acknowledgments}
This study was partly supported by JSPS KAKENHI No.~JP23KJ0756 (Y.K.), JSPS Overseas Challenge Program for Young Researchers (Y.K.), and Deutsche Forschungsgemeinschaft (DFG) -- project-ID 318763901 (SFB1294). 

Y.K. thanks fruitful discussions with H.~Ishii, K.~Hayashi, H.~Takagi, M.~Tarama, Y.~Izumida, and H.~Kori. 

The authors used ChatGPT 5.5 and Claude Opus 4.8 to assist with reviewing and developing parts of the code, and OpenAI Prism (ChatGPT 5.2) to assist with manuscript revision. All scientific content, interpretations, code, and manuscript text were reviewed and verified by the authors. 
\end{acknowledgments}

\section*{Data availability}
The data that support the findings of this article are openly available \cite{github}. 

\appendix

\cleardoublepage

\section{Details of candidate models}
\label{sec:app_model_detail}
Here, we describe the statistical quantities for each candidate models. We define the MSD and VACF for continuous models by
\begin{equation}
    \label{eq:msd}
    \hat{m}(\tau) \coloneqq \langle \| {\bm r}(\tau)-{\bm r}(0)\|^2 \rangle, 
\end{equation}
and
\begin{equation}
    \label{eq:vacf}
    C_v(\tau) \coloneqq \langle \bm v(t+\tau)\cdot \bm v(t) \rangle, 
\end{equation}
respectively.

\subsection{Brownian motion}
The MSD for the BM model is
\begin{equation}
    \label{eq:BM_msd}
    m(\tau) = 4D\tau, 
\end{equation}
in two spatial dimensions. 

\subsection{Integrated OU process}
In the stationary state, the MSD of the OU model~\eqref{eq:OU} is given by~\cite{ebeling2005statistical}
\begin{equation}
    \label{eq:OU_msd}
    \hat m(\tau) = \frac{4D}{\gamma^3}(e^{-\gamma \tau} + \gamma \tau -1), 
\end{equation}
while the VACF is
\begin{equation}
    C_v(\tau) = \frac{2D}{\gamma} e^{-\gamma \tau}. 
    \label{eq:OU_vacf}
\end{equation}
These two quantities show that the OU model contains a single characteristic time scale, $\gamma^{-1}$. For $\tau \ll \gamma^{-1}$, the velocity remains correlated over the observation interval, so the displacement is approximately proportional to $\tau$ and the MSD scales ballistically as $\tau^2$. On the other hand, for $\tau \gg \gamma^{-1}$, the velocity autocorrelation has decayed and the motion becomes diffusive with an effective diffusion coefficient $D/\gamma^2$.

\subsection{Mexican-hat model}
Because of the nonlinear drift term, closed-form MSD and VACF expressions are not available in the Mexican-hat model~\eqref{eq:Mhat}. Nevertheless, the stationary velocity distribution can be obtained in closed form from the corresponding Fokker--Planck equation~\cite{erdmann2000brownian}:
\begin{equation}
    \label{eq:Mhat_stat_dist}
    p_{\mathrm{s}}({\bm v}) = \frac{1}{C} \exp \left(- \frac{\gamma}{4D} \| \bm v \|^4 + \frac{\gamma v_r^2}{2D} \| \bm v \|^2 \right) ,
\end{equation}
where the normalization constant $C$ is given as
\begin{align}
    C = \pi^{3/2} \sqrt{\frac{D}{\gamma}} \exp \left( \frac{\gamma v_r^4}{4 D} \right) \left[ 1 + \mathrm{erf}\left( \frac{v_r^2}{2}\sqrt{\frac{\gamma}{D}} \right) \right],
\end{align}
with the error function defined by
\begin{equation}
    \mathrm{erf} (z) \coloneqq \frac{2}{\sqrt{\pi}}\int_{0}^{z} e^{-t^2} dt.
\end{equation}
For $v_r>0$, this distribution is maximal on the circle $\|\bm v\|=v_r$ and is suppressed near zero and at very large speeds. 

\subsection{Memory kernel model}
\label{app:der_memory}
By introducing an auxiliary variable 
\begin{equation}
    \label{eq:fly_def_w}
    {\bm w} \coloneqq \int_{-\infty}^{t} dt' e^{-a (t-t')} \boldsymbol{v}(t'),
\end{equation}
we can rewrite Eq.~\eqref{eq:fly} as
\begin{subequations}
    \label{eq:fly_2}
    \begin{align}
        \dot{\bm r} &= {\bm v}, \\
        \dot{\bm v} &= -\gamma \bm v + A \bm{w} + \sqrt{2D} {\bm \xi}(t), \label{eq:fly_v_new} \\
        \dot{\bm w} &= -a \bm{w} + \bm{v}. \label{eq:fly_w}
    \end{align}
\end{subequations}
The stability condition~\eqref{eq:fly_assump} ensures that the origin $[(\bm{v}, \bm{w}) = (\bm{0}, \bm{0})]$ is stable for the deterministic part of the subsystem defined by Eqs.~\eqref{eq:fly_v_new} and~\eqref{eq:fly_w}.

In the stationary state, the VACF for the memory kernel model is
\begin{align}
    C_v(\tau) = 2D \left(C_1 e^{-d_1 |\tau|} + C_2 e^{-d_2 |\tau|} \right) \, , \label{eq:fly_VACF}
\end{align}
where 
\begin{subequations}
\label{eq:def_diBi}
    \begin{align}
        d_{1,2} &\coloneqq \frac{a + \gamma \mp \sqrt{4A + (a-\gamma)^2}}{2} \, ,  \label{eq:fly_d12} \\
        C_1 &\coloneqq \frac{a^2 - d_1^2}{d_1(d_2^2 - d_1^2)} \, , \quad {\rm and} \quad 
        C_2 \coloneqq \frac{d_2^2 - a^2}{d_2(d_2^2 - d_1^2)} \, .
    \end{align}
\end{subequations}
The MSD in the stationary state is
\begin{align}
    \hat m(\tau)
    & = 4D \sum_{k=1,2} \frac{C_k}{d_k^2} \left( e^{-d_k \tau} + d_k \tau - 1 \right) \, \label{eq:fly_msd}.
\end{align}
We provide the derivations of Eq.~\eqref{eq:fly_VACF} and \eqref{eq:fly_msd} in \ref*{SI:sec:fly_vacf} of Ref.~\cite{SI}.  

\section{Importance sampling with Gaussian mixture proposal}
\label{sec:app_method_detail}

\subsection{Method}
To compute the log Bayes factors in Eq.~\eqref{eq:lbf} and hence to perform model comparison, we evaluate the marginal likelihood
\begin{align}
    \label{eq:ml_simple}
    Z
    &\coloneqq
    \int d{\bm \theta}\,
    p(\mathcal{D}|{\bm \theta})\, p({\bm \theta}), 
\end{align}
which is Eq.~\eqref{eq:ml} written for a fixed candidate model, with the model label and index suppressed. Since this integral is typically not available analytically, we approximate it by importance sampling (IS). For a proposal distribution $q(\bm \theta)$ whose support contains the posterior support, we write
\begin{align}
    Z
    &=
    \int d{\bm \theta}\, q({\bm \theta})\,
    \frac{p(\mathcal{D}|{\bm \theta})p({\bm \theta})}{q({\bm \theta})}
    \notag \\
    &\approx
    \frac{1}{N}\sum_{n=1}^{N} w_n, \label{eq:IS_intro}
\end{align}
where
\begin{equation}
    w_n\coloneqq\frac{p(\mathcal{D}|\bm \theta^{(n)})p(\bm \theta^{(n)})}{q(\bm \theta^{(n)})},
    \label{eq:IS-estimator}
\end{equation}
denotes the importance weight and $\{\bm \theta^{(n)}\}_{n=1}^{N}$ are $N$ independent samples drawn from $q(\bm \theta)$. 

The efficiency of the IS estimator in Eq.~\eqref{eq:IS_intro} depends strongly on how well the proposal overlaps with the posterior distribution $p({\bm \theta}|\mathcal{D})$. To better capture the structure of the posterior, we use a mixture proposal~\cite{owen2000safe,elvira2016heretical,elvira2019generalized}. In particular, to cover multiple posterior modes while retaining a simple sampling procedure, we use a Gaussian mixture model (GMM):
\begin{align}
    \label{eq:proposal_gmm}
    q_{\mathrm{GMM}}(\bm \theta)
    &=
    \sum_{j=1}^{K}\pi_j\,
    \mathcal N\bigl(\bm \theta;\,\tilde{\bm{\theta}}_j,\,\Lambda_j \bigr),
\end{align}
where $\{\tilde{\bm{\theta}}_j\}_{j=1}^{K}$ are local maxima of the following log unnormalized posterior density:
\begin{equation}
    \rho(\bm \theta) \coloneqq \ln p(\mathcal{D}|{\bm \theta}) + \ln p({\bm \theta}),
\end{equation}
$K$ is the number of detected modes, $\Lambda_j$ is a local covariance matrix estimated from a Laplace approximation of $\rho$ at $\tilde{\bm{\theta}}_j$, and $\pi_j > 0$ is a mixture weight satisfying $\sum_{j=1}^{K} \pi_j = 1$. We optimize $\rho$ from multiple initial points to detect multiple modes. 
Further details of the mode search, covariance construction, treatment of non-negative-definite Hessians, and stable evaluation of the sum in Eq.~\eqref{eq:IS_intro} are given in \ref*{SI:sec:detail_IS} of Ref.~\cite{SI}.

To further improve robustness, we combine the GMM proposal with the prior distribution, yielding the defensive mixture proposal~\cite{hesterberg1995weighted}:
\begin{align}
    q(\bm \theta)
    &= (1-\omega)\,q_{\mathrm{GMM}}(\bm \theta) + \omega \,p(\bm \theta),
    \label{eq:proposal-mixture-with-prior}
\end{align}
where we set $\omega = 0.5$.
This construction guarantees global support wherever $p(\bm \theta)>0$ and reduces sensitivity to posterior mass missed by the GMM component. We use the proposal in Eq.~\eqref{eq:proposal-mixture-with-prior} to compute the IS estimator in Eq.~\eqref{eq:IS_intro}. 


We assess the efficiency of importance sampling by monitoring the effective sample size (ESS)~\cite{kong1992note,kong1994sequential, liu2001monte}, defined as
\begin{equation}
    \label{eq:ess}
    \mathrm{ESS} \coloneqq \frac{\left[\sum_{n=1}^{N} w(\bm \theta^{(n)})\right]^2}{\sum_{n=1}^{N} w(\bm \theta^{(n)})^2}. 
\end{equation}
If the proposal $q(\bm \theta)$ is identical to the posterior $p(\bm \theta | \mathcal{D})$, all weights $w_1, \ldots, w_N$ are equal and $\mathrm{ESS}=N$. In contrast, if only $k$ samples have non-negligible weights while the remaining weights are nearly zero, then $\mathrm{ESS}\approx k$.
We use the ESS as a diagnostic of proposal quality and to guide the choice of numerical hyperparameters, such as $\omega$ in Eq.~\eqref{eq:proposal-mixture-with-prior}. 

Finally, to improve numerical stability, we evaluate the weights in Eq.~\eqref{eq:IS-estimator} in log-space,
\begin{equation}
    \ln w_n=\log p(\mathcal{D}|\bm\theta^{(n)})+\ln p(\bm\theta^{(n)})-\ln q(\bm\theta^{(n)}),
\end{equation}
and compute the sum in Eq.~\eqref{eq:IS_intro} using the log-sum-exp technique to avoid numerical overflow~\cite{blanchard2021accurately}.

\subsection{Robustness of importance sampling}

\begin{figure}
    \centering
    \includegraphics[width=1.\linewidth]{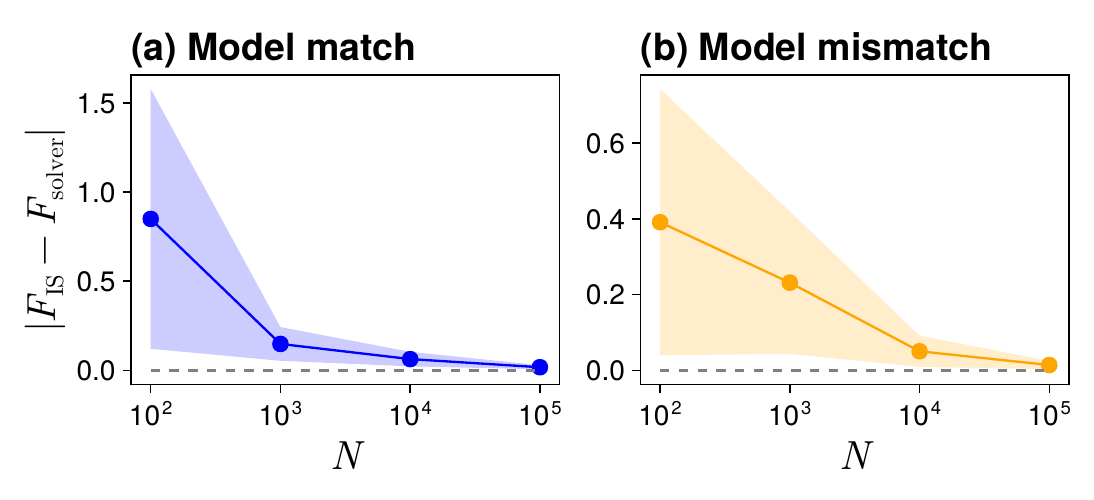}
    \caption{Accuracy of the importance-sampling estimate. We compare the negative logarithm of the marginal likelihood [Eq.~\eqref{eq:bfe}] obtained by direct numerical integration, $F_{\mathrm{solver}}$, with that obtained by importance sampling, $F_{\mathrm{IS}}$, and plot the absolute difference $|F_{\mathrm{solver}}-F_{\mathrm{IS}}|$ as a function of the number of samples $N$. Trajectories are generated with $T=10$ and $\Delta t=0.1$: from a Mexican-hat model with $\gamma=D=1$ and $v_r=2$ in panel (a), and from an OU model with $\gamma=D=1$ in panel (b). In both panels, the Mexican-hat model is used as the inference model with test priors $p(\gamma)=p(D)=p(v_r)=\mathcal{L}(0.1,5)$. For each panel, we generate 20 trajectories with different random seeds and calculate the marginal likelihood for each trajectory. Points and shaded regions indicate the mean and standard deviation across the $20$ trajectories, respectively. In both the model-match case [panel (a)] and the model-mismatch case [panel (b)], the error decreases as $N$ increases.}
    \label{fig:verify_IS}
\end{figure}

We first validate the accuracy of the importance-sampling estimator~\eqref{eq:IS-estimator} by comparing it with the ``true'' marginal likelihood obtained from direct numerical integration of Eq.~\eqref{eq:ml}. For numerical convenience, we work with the negative logarithm of the marginal likelihood,
\begin{align}
    \label{eq:bfe}
    F_i \coloneqq - \ln Z_i. 
\end{align} 
For the direct integration, we use a Julia implementation of a multidimensional adaptive integration scheme~\cite{hcubature}, which is based on Ref.~\cite{genz1980remarks}. The results are shown in Fig.~\ref{fig:verify_IS}. We consider both a model-match case, in which the data-generating model coincides with the inference model [panel (a)], and a model-mismatch case, in which the two models differ [panel (b)]. For this validation, we use shorter trajectories and narrower priors than in the main analyses to keep the direct numerical integration stable and computationally feasible. 
Figure~\ref{fig:verify_IS} shows that the discrepancy between the direct solver and the importance-sampling estimate decreases toward zero as the number of samples $N$ increases. Based on this convergence, we use $N = 10^5$ samples in the subsequent analyses.

\section{Details in likelihood calculation from positional increments}
\subsection{Limitations of secant-velocity approximations}
\label{sec:app_lim_secant}
When constructing a likelihood for second-order models from position-only observations, a tempting but generally incorrect shortcut is to estimate the velocity by finite differences and substitute it into a likelihood written for the instantaneous velocity. In other words, we cannot simply replace the unobserved instantaneous velocity ${\bm v}_i$ by the secant velocity $\bm V_i$ in Eq.~\eqref{eq:secant-velocity}, because ${\bm v}_i$ and $\bm V_i$ are different stochastic processes for a finite sampling interval $\Delta t$. 

For the second-order models (except for the memory kernel model), the sampled instantaneous velocities ${\bm v}_i$ form a Markov process, whereas the secant velocities ${\bm V}_i$ generally do not. 
Indeed,
\begin{equation}
    \bm V_i = \frac{1}{\Delta t}\int_{t_i}^{t_{i+1}} \bm v(s)\, ds, 
    \label{eq:secant-average}
\end{equation}
hence $\bm V_i$ is an average over the interval $[t_i,t_{i+1}]$, and thus loses information about the endpoint velocity that controls the subsequent dynamics. The distinction between $\bm V_i$ and $\bm v_i$ is also clear from the component-wise covariance
\begin{align}
    & \langle V_i^{(\alpha)} V_j^{(\alpha)} \rangle
    \notag \\ 
    = & \frac{1}{\Delta t^2}
    \int_{t_i}^{t_{i+1}} ds \int_{t_j}^{t_{j+1}} ds'\,
    \langle v^{(\alpha)}(s) \, v^{(\alpha)}(s') \rangle.
    \label{eq:secant-covariance}
\end{align}
with $\alpha \in \{x,y\}$. This generally differs from the covariance of the discrete instantaneous velocity, $\langle v_i^{(\alpha)}v_j^{(\alpha)}\rangle$. Therefore, replacing $\bm v_i$ by $\bm V_i$ in an instantaneous-velocity likelihood gives a misspecified likelihood and can bias parameter estimates~\cite{ferretti2020building,ferretti2022renormalization,albrecht2024inferring}. This motivates constructing the likelihood directly from position increments rather than from reconstructed velocities.


\subsection{Linear Gaussian models}
\label{sec:method_likeli_linear}
For the linear Gaussian candidates, the positional increments remain jointly Gaussian after the latent velocities and auxiliary hidden variables are integrated out~\cite{lysy2016model}. Therefore, for each Cartesian component $\alpha \in \{x,y\}$, the increment vector $\Delta \bm r^{(\alpha)}$ in Eq.~\eqref{eq:increment_vec} follows a zero-mean multivariate Gaussian distribution,
\begin{equation}
    \Delta \bm r^{(\alpha)} \sim \mathcal{N}(\bm 0, \Sigma).
\end{equation}
The mean is zero because the stationary dynamics have no systematic drift. 
As in Eq.~\eqref{eq:secant-covariance}, the covariance matrix $\Sigma$ is obtained by integrating the component-wise VACF over the two sampling intervals, i.e., 
\begin{align}
    \Sigma_{ij} & \coloneqq \left\langle \Delta r_i^{(\alpha)} \Delta r_j^{(\alpha)} \right\rangle \notag \\
    & = \int_{t_i}^{t_{i+1}} ds \int_{t_j}^{t_{j+1}} ds'\,
    \langle v^{(\alpha)}(s) \, v^{(\alpha)}(s') \rangle.
    \label{eq:position-covariance}
\end{align}
For the OU model, substituting the VACF in Eq.~\eqref{eq:OU_vacf} into Eq.~\eqref{eq:position-covariance} gives 
\begin{align}
    \label{eq:cov_OU}
    \Sigma_{ij} = 
    \begin{cases}
        \displaystyle \frac{2D}{\gamma^3} (\gamma \Delta t + e^{-\gamma \Delta t} - 1) & i=j, \\
        \displaystyle \frac{D}{\gamma^3} e^{-\gamma \Delta t (|i-j|-1)} (1-e^{-\gamma \Delta t})^2 & i \neq j,
    \end{cases}
\end{align} 
We note that the component-wise VACF [i.e., $\langle v^{(\alpha)}(s) \, v^{(\alpha)}(s') \rangle$ in Eq.~\eqref{eq:position-covariance}] is half of the vector VACF in Eq.~\eqref{eq:OU_vacf}. 
Similarly, for the memory kernel model with the VACF in Eq.~\eqref{eq:fly_VACF}, the covariance is 
\begin{equation}
    \label{eq:cov_MK}
    \Sigma_{ij} = 
    \begin{cases}
        \displaystyle \sum_{k = 1,2} \frac{2D C_k}{d_k^2} (d_k \Delta t + e^{-d_k \Delta t} - 1) & i=j, \\
        \displaystyle \sum_{k = 1,2} \frac{D C_k}{d_k^2} e^{-d_k \Delta t (|i-j|-1)} (1-e^{-d_k \Delta t})^2 & i \neq j.
    \end{cases}
\end{equation}

For a fixed initial position $\bm r_0$, the likelihood for the observed trajectory is equivalently the joint density of the positional increments. Since the two Cartesian components are independent and have the same covariance matrix, this positional likelihood becomes Eq.~\eqref{eq:likeli_lin_gauss}. 

Because these covariance matrices [Eq.~\eqref{eq:cov_OU} and \eqref{eq:cov_MK}] are symmetric Toeplitz, i.e., $\Sigma_{ij}$ depends only on $|i-j|$, the likelihood can be evaluated efficiently using the recursive Levinson--Durbin algorithm detailed in \ref*{sec:toeplitz-recursion} in Ref.~\cite{SI}.

\subsection{Nonlinear models}
\label{sec:method_likeli_nonlinear}
For nonlinear models, an exact closed-form increment likelihood is, in general, not available. We therefore use the transformed-Gaussian approach proposed in Ref.~\cite{albrecht2024inferring} to derive the approximate likelihood. Below, we briefly review their approach. 

Consider a second-order model of the form
\begin{subequations}
    \begin{align}
        \dot{\bm r} &= \bm v, \\
        \dot{\bm v} &= \bm f (\bm v; \bm \theta) + \sqrt{2D} \bm \xi (t), \label{eq:general_v}
    \end{align}
\end{subequations}
with a nonlinear drift term $\bm f(\bm v;\bm \theta)$. For sufficiently small sampling interval $\Delta t$, integrating Eq.~\eqref{eq:general_v} over one observation interval gives an approximate discrete-time equation for the secant velocities~\eqref{eq:secant-velocity}. In the notation of Ref.~\cite{albrecht2024inferring}, this leads to the transformed variable
\begin{align}
    \bm Q_i &\coloneqq \bm V_{i+1}-\bm V_i- \frac{\Delta t}{2}\, \left[\bm f(\bm V_i)+\bm f(\bm V_{i+1})\right],  
\end{align}
for $i = 0, \ldots, M-2$. For each Cartesian component $\alpha$, the transformed sequence 
\begin{equation}
    \label{eq:def_nlin_Qalpha}
    \bm Q^{(\alpha)} \coloneqq (Q_0^{(\alpha)},\dots,Q_{M-2}^{(\alpha)})^{\top}, 
\end{equation} 
is approximated as Gaussian,
\begin{equation}
    \bm Q^{(\alpha)} \sim \mathcal N\!\left(\bm 0, 2D\,\bm Z\right),
\end{equation}
where $\bm Z$ is the following symmetric tridiagonal covariance matrix of order $M-1$: 
\begin{equation}
    \label{eq:def_nlin_Z}
    \bm Z = \frac{\Delta t}{6}
    \begin{pmatrix}
        4 & 1 & 0 & \cdots & 0 \\
        1 & 4 & 1 & \ddots & \vdots \\
        0 & 1 & 4 & \ddots & 0 \\
        \vdots & \ddots & \ddots & \ddots & 1 \\
        0 & \cdots & 0 & 1 & 4
    \end{pmatrix},
\end{equation}
which arises from finite-time averaging of the noise over the sampling interval~\cite{albrecht2024inferring}. Transforming variables from $\{\bm Q_i\}$ back to $\{\bm V_i\}$ gives the approximate likelihood for the secant-velocity sequence, including the associated Jacobian correction:
\begin{align}
    & p(\bm V_0,\dots,\bm V_{M-1} | \bm \theta) \notag \\
    \approx\, & p(\bm V_0 | \bm \theta)
    \prod_{\alpha = x,y}
    \mathcal N\!\left(\bm Q^{(\alpha)};\bm 0, 2D\,\bm Z\right) \notag \\
    & \qquad \times
    \prod_{i=1}^{M-1}
    \left|\det\!\left(I-\frac{\Delta t}{2}J[\bm f](\bm V_i)\right)\right|. \label{eq:likeli_nonlin_secant}
\end{align}
We approximate $p(\bm V_0 | \bm \theta)$ in practice by the stationary distribution of the instantaneous velocity~\cite{albrecht2024inferring}. Finally, for fixed initial position $\bm r_0$, the likelihood of the observed positions is obtained from the change of variables between increments and secant velocities:
\begin{equation}
    \label{eq:likeli_nonlin_approx}
    p(\mathcal{D} | \bm \theta) = (\Delta t)^{-2M}\, p(\bm V_0,\dots,\bm V_{M-1} | \bm \theta). 
\end{equation}
This gives a tractable approximate likelihood for nonlinear candidate models.

\section{Experimental setup and data processing}
\label{app:exp_detail}
We investigate data of three strains of
\textit{Dictyostelium discoideum}: \textit{Dictyostelium discoideum} DdB, DdB NF1 KO, and AX2. The results for DdB are reported in the main manuscript while the remaining results can be found in the Supplementary Information.

The cells were imaged after non-axenic cultivation with \textit{Klebsiella aerogenes}; AX2 cells were additionally grown axenically in HL5 medium.
For non-axenic cultivation, cells were maintained for 3--4 days at \qty{22}{\degreeCelsius} in \qty{10}{\centi\meter} dishes containing \qty{9.5}{\milli\liter} S{\o}rensen's buffer and \qty{0.5}{\milli\liter} bacterial feeding suspension at $\mathrm{OD}_{600}=2$.
For DdB NF1 KO cultures, G418 was added 24~h after inoculation to a final concentration of \qty{5}{\micro\gram\per\milli\liter}.
Axenic AX2 cultures were maintained in \qty{10}{\milli\liter} HL5 medium supplemented with \qty{10}{\micro\liter} PenStrep.
The \textit{K.~aerogenes} feeding suspension was prepared by growth in LB medium, followed by centrifugation, washing, and resuspension in S{\o}rensen's buffer to $\mathrm{OD}_{600}=2$.
Cells were harvested at confluence before overgrowth.
Residual bacteria were removed from non-axenic samples by a 1-h incubation on a shaking table followed by three washing steps in S{\o}rensen's buffer at $300\times g$ for 5~min.
The washed cells were diluted to minimize cell--cell contacts, transferred to an ibidi $\mu$-Slide 8 Well$^{\mathrm{high}}$ Grid-500, and allowed to settle for 25~min before imaging.

Bright-field time-lapse recordings were acquired using an HTCS wide-field microscope equipped with an XM10 monochrome CCD camera.
A motorized stage sequentially revisited six fixed positions within the slide grid.
At each position, images were acquired with an exposure time of \qty{20}{\milli\second} at an effective frame interval of \qty{5}{\second} for \qty{2}{\hour}, yielding 1440 frames per image sequence.

Stage-induced image displacement was corrected using the fixed grid dots of the slide as fiducial markers.
For each frame, the mean translational displacement relative to the reference frame was calculated from at least 40, and typically approximately 400, valid grid-marker trajectories and subtracted from the image coordinates.
The registered image sequences were cropped to exclude the grid-marker regions.
Cells were subsequently identified by threshold-based binarization and a modified MATLAB active-contour segmentation procedure based on code originally developed by Meghan K. Driscoll, made available to our group, and further adapted for the image-analysis workflow used here; the resulting centroid positions were linked across consecutive frames to obtain single-cell trajectories.
Tracks were terminated when cells left the field of view or came into contact with neighboring cells and were reinitialized after individual cells became distinguishable again.
All trajectories were visually inspected and tracks affected by segmentation errors, unresolved contacts, or other tracking artifacts were manually excluded from the statistical analysis.

When the tracking algorithm temporarily fails to detect a cell, the corresponding position is recorded as a ``NaN'' value. For trajectories containing such missing entries, we split the raw trajectory at the missing positions and use the longest contiguous segment without missing values as the trajectory for that cell. To exclude nearly immobile cells, we remove trajectories whose maximum displacement from the initial position is less than or equal to $\qty{2}{\micro\meter}$. 

The datasets reported in the Supplementary Information, were subjected to the same filtering as described in Sec.~\ref{sec:experiment} for the DdB dataset: only trajectories with at least 257 recorded positions were retained. After filtering, the datasets contain 464 trajectories for AX2 cells grown with bacteria, 566 for AX2 cells grown without bacteria, and 670 for NF1 knockout cells. 

\section{Estimated parameters from experimental trajectories}
\label{app:exp_paraest}

\begin{figure}[h]
    \centering
    \includegraphics[width=1.\linewidth]{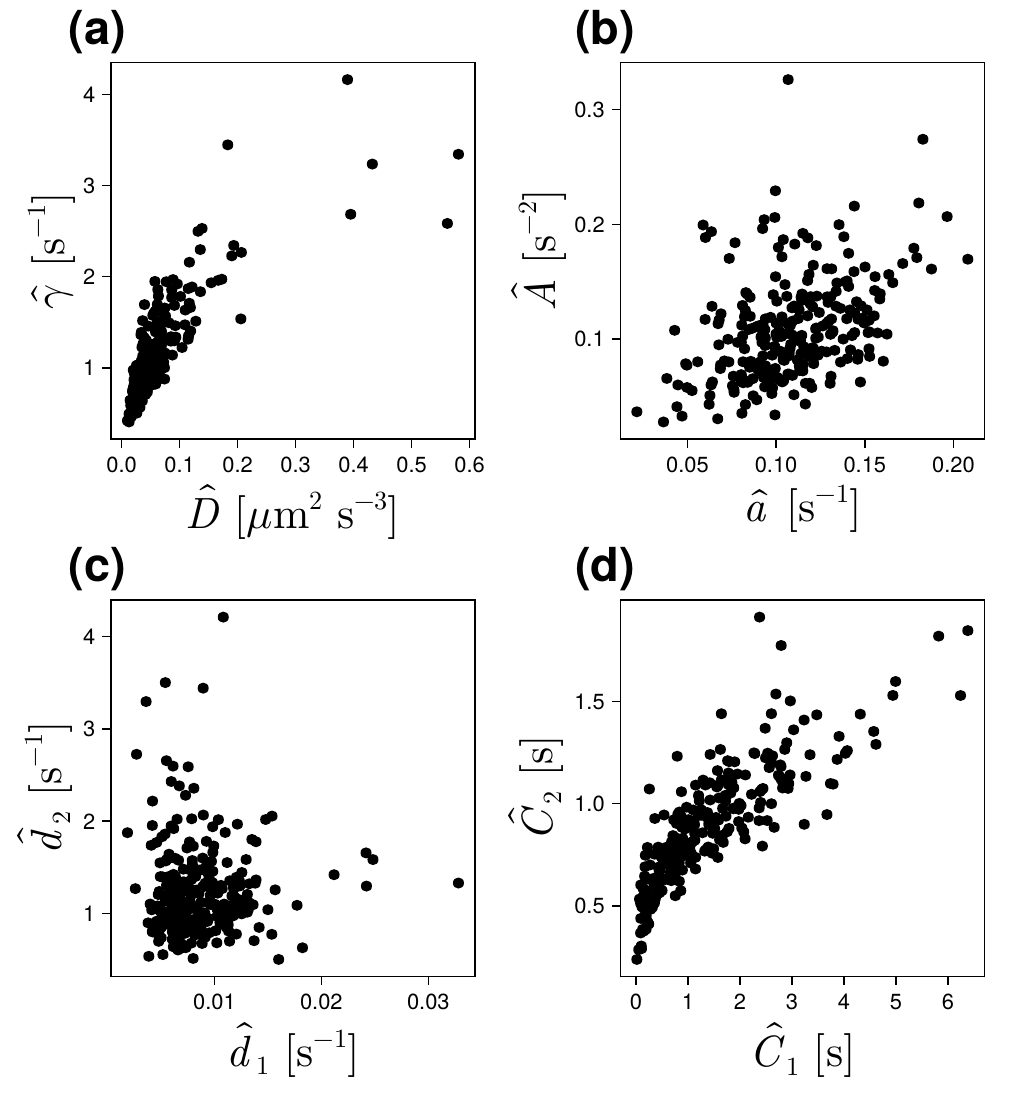}
    \caption{Parameter estimates for the memory kernel model. Results are shown for the $261$ DdB trajectories at sampling interval $\Delta t = \qty{5}{\second}$. Panels (a) and (b) show scatter plots of the MAP estimates for the original parameters $(\hat{D}, \hat{\gamma}, \hat{a}, \hat{A})$. Panels (c) and (d) show the corresponding time scales $(\hat{d}_1,\hat{d}_2)$ and their coefficients $(\hat{C}_1, \hat{C}_2)$, respectively.}
    \label{fig:DdB_comp_5sec}
\end{figure}

Here, we present the inferred parameters of the memory kernel model for the DdB trajectories sampled at the original interval, $\Delta t = \qty{5}{\second}$. Figure~\ref{fig:DdB_comp_5sec} shows scatter plots of the MAP estimates~\eqref{eq:map} of the original model parameters, $(\hat{D},\hat{\gamma},\hat{a},\hat{A})$, in panels (a) and (b), the corresponding characteristic time scales, $(\hat{d}_1,\hat{d}_2)$, in panel (c), and their coefficients, $(\hat{C}_1,\hat{C}_2)$, in panel (d). These MAP estimates are also used to compute the model predictions of the MSD and VACF shown in Fig.~\ref{fig:DdB_comp_5sec_prediction}. The inferred parameters exhibit substantial heterogeneity across trajectories.

\newpage
\bibliography{motility}

\end{document}


\title{Supplementary Information for ``Bayesian comparison of Langevin dynamics for cell motility from positional observation''}

    \author{Yusuke Kato}
    \email{yusukeka@umich.edu}
    \affiliation{Gilbert S.~Omenn Department of Computational Medicine and Bioinformatics, University of Michigan, Ann Arbor, 48109-2218, MI, USA}
    \affiliation{Department of Complexity Science and Engineering, Graduate School of Frontier Sciences, The University of Tokyo, Kashiwa, Chiba 277-8561, Japan}
    
    \author{Jan Albrecht}
    \affiliation{Institute of Physics and Astronomy, University of Potsdam, 14476 Potsdam, Germany}
    
    \author{Ted Moldenhawer}
    \affiliation{Institute of Physics and Astronomy, University of Potsdam, 14476 Potsdam, Germany}
    
    \author{Robert Gro\ss mann}
    \affiliation{Institute of Physics and Astronomy, University of Potsdam, 14476 Potsdam, Germany}
    
    \author{Carsten Beta}
    \affiliation{Institute of Physics and Astronomy, University of Potsdam, 14476 Potsdam, Germany}
    \affiliation{Nano Life Science Institute (WPI-NanoLSI), Kanazawa University, Kakuma-machi, Kanazawa 920-1192, Japan}
	
\maketitle
\setcounter{page}{1}
\setcounter{figure}{0}

\tableofcontents

\section{Derivations of VACF and MSD for the memory kernel model}
\label{SI:sec:fly_vacf}

We work on the memory kernel model with auxiliary variable $\bm w$, i.e., Eq.~\eqrefstar{eq:fly_2} in the main text. Applying the Fourier transform, defined here as 
\begin{equation}
    \hat{\bm{x}}(q) \coloneqq \int_{-\infty}^{\infty} dt\, \bm{x}(t) e^{-i qt} \, , 
\end{equation}
yields
\begin{align}
    iq \hat{\bm r} &= \hat{\bm{v}}, \\
    iq \hat{\bm v} &= -\gamma \hat{\bm{v}} + A \hat{\bm{w}} + \sqrt{2D} \hat{\bm{\xi}}, \label{eq:fly_v_f} \\
    iq \hat{\bm w} &= -a \hat{\bm{w}} + \hat{\bm{v}}. \label{eq:fly_w_f}
\end{align}
Eqs.~\eqref{eq:fly_v_f} and~\eqref{eq:fly_w_f} imply that 
\begin{equation}
    \hat{\bm{v}} = K(q) \hat{\bm{\xi}}, \label{eq:fly_hatv_hatxi}
\end{equation}
where
\begin{equation}
    K(q) \coloneqq \frac{\sqrt{2D}}{\gamma + iq - \frac{A}{a + iq}}. 
\end{equation}

Since $\xi^{(x)}$ and $\xi^{(y)}$ are uncorrelated, each coordinate $\alpha \in \{x,y\}$ can be treated independently. 
The autocorrelation of the Fourier-transformed noise is 
\begin{align}
    \langle \hat{\xi}^{(\alpha)} (q) \, \hat{\xi}^{(\alpha)} (q') \rangle 
    &= \int dt\, \int dt'\, \delta(t-t') e^{-i qt} e^{-i q't'} \notag \\
    &= 2 \pi \delta(q+q'), 
\end{align}
which implies that
\begin{align}
    \langle \hat{v}^{(\alpha)} (q) \, \hat{v}^{(\alpha)} (q') \rangle 
    &= 2 \pi K(q) K(q') \delta(q+q').
\end{align}
Applying the inverse Fourier transform gives
\begin{align}
    \langle v^{(\alpha)} (t) \, v^{(\alpha)} (t') \rangle 
    &= \frac{1}{2 \pi} \int dq\, K(q) K(-q) e^{i q(t-t')}.
\end{align}
Noting that
\begin{align}
    K(q) K(-q) 
    &= \frac{2D (a^2 + q^2)}{q^4 + [(a^2 + \gamma^2) + 2A] q^2 + (A - a\gamma)^2} \notag \\
    &= 2D \left( \frac{d_1 C_1}{q^2 + d_1^2} + \frac{d_2 C_2}{q^2 + d_2^2} \right),
\end{align}
and using the formula obtained from Jordan's lemma, 
\begin{equation}
    \int_{-\infty}^{\infty} dq\, \frac{e^{i q \tau}}{q^2 + k^2} = \frac{\pi}{k} e^{-k |\tau|}, 
\end{equation}
for $k>0$, we obtain
\begin{align}
    & \langle v^{(\alpha)} (t) \, v^{(\alpha)} (t') \rangle \notag \\
    = & \frac{D}{\pi} \left[ d_1 C_1 \int dq\, \frac{e^{i q(t-t')}}{q^2 + d_1^2} + d_2 C_2 \int dq\, \frac{e^{i q(t-t')}}{q^2 + d_2^2} \right]  \notag \\
    = & D \left(C_1 e^{-d_1 |t-t'|} + C_2 e^{-d_2 |t-t'|} \right).
    \label{eq:fly_velcor_each}
\end{align}
Summing the autocorrelation in Eq.~\eqref{eq:fly_velcor_each} over the two coordinates gives Eq.~\eqrefstar{eq:fly_VACF}.
We can also derive the crossover time of this double-exponential VACF as the time at which the two exponential contributions are equal, $C_1 e^{-d_1 |\tau|} = C_2 e^{-d_2 |\tau|}$. This condition gives
\begin{equation}
\label{eq:fly_cross_time}
    \tau_{\mathrm{cross}} \coloneqq \frac{\ln (C_2 / C_1)}{d_2 - d_1}.  
\end{equation}
This quantity provides another characteristic timescale of the memory kernel model~\eqrefstar{eq:fly}. 


Using Eq.~\eqref{eq:fly_velcor_each}, the MSD of the stationary process is derived as
\begin{align}
    \hat m(\tau)
    & = \sum_{\alpha = x,y} \int_0^{\tau} \int_0^{\tau} \left \langle v^{(\alpha)} (s) \, v^{(\alpha)} (s') \right \rangle ds ds' \notag \\
    & = 4D \sum_{k=1,2} \frac{C_k}{d_k^2} \left( e^{-d_k \tau} + d_k \tau - 1 \right) \, \label{eq:fly_msd}.
\end{align}

\clearpage
\section{Levinson-Durbin Algorithm}
\label{sec:toeplitz-recursion}
When the covariance matrix is an $n$-dimensional symmetric Toeplitz matrix, the multivariate normal density 
can be evaluated recursively with $O(n^2)$ computational cost using the Levinson--Durbin algorithm~\cite{sayed2001survey}. 

Let $C_n$ be a positive definite symmetric Toeplitz matrix of order $n$: 
\begin{equation}
    C_n \coloneqq 
    \begin{pmatrix}
        c_0 & c_1 & \cdots & c_{n-1} \\
        c_1 & c_0 & & c_{n-2} \\
        \vdots & & \ddots & \vdots \\
        c_{n-1} & c_{n-2} & \cdots & c_0 
    \end{pmatrix}
    .
\end{equation}
For a given vector $\bm x_n=(x_1, \dots, x_n)^\top$, we consider the solution $\bm y_n$ of the following linear equation
\begin{equation}
    \label{eq:obj_eq_LD}
    C_n \bm y_n = \bm x_n. 
\end{equation}
We denote by $\bm y_k=(y_k^{(1)}, \dots, y_k^{(k)})^\top$ the solution of the leading principal system
\begin{equation}
\label{eq:def_ck_eq}
    C_k \bm y_k = \bm x_k,
\end{equation}
with $\bm x_k = (x_1, \dots, x_k)^\top$ for $1 \leq k \leq n$. 
We solve $C_{k+1}\bm y_{k+1}=\bm x_{k+1}$ by rewriting
\begin{equation}
    \label{eq:def_ck+1_eq}
    \begin{pmatrix}
        C_k & \bm r_k \\
        \bm r_k^\top & c_0
    \end{pmatrix}
    \begin{pmatrix}
        \bm u \\
        \alpha
    \end{pmatrix}
    = 
    \begin{pmatrix}
        \bm x_k \\
        x_{k+1}
    \end{pmatrix}
    ,
\end{equation}
with 
\begin{align}
    \bm r_k & \coloneqq (c_k,c_{k-1},\dots,c_1)^\top, \\
    \bm u & \coloneqq (y_{k+1}^{(1)}, y_{k+1}^{(2)},\dots, y_{k+1}^{(k)})^\top, 
\end{align}
and
\begin{equation}
    \alpha \coloneqq y_{k+1}^{(k+1)}.    
\end{equation}    
Equation~\eqref{eq:def_ck+1_eq} implies that
\begin{equation}
    \bm u = C_k^{-1}(\bm x_k-\bm r_k \alpha) = \bm y_k - \alpha C_k^{-1} \bm r_k,
\end{equation}
and
\begin{equation}
    \alpha
    =
    \frac{x_{k+1}-\bm r_k^\top \bm y_k}{c_0 - \bm r_k^\top C_k^{-1} \bm r_k}.
\end{equation}
By setting
\begin{equation}
\label{eq:def_ak}
    \bm a_k \coloneqq C_k^{-1}\bm r_k, 
\end{equation}
and 
\begin{equation}
    s_k \coloneqq c_0-\bm r_k^\top \bm a_k, 
\end{equation}
we finally get
\begin{equation}
    \label{eq:y-update-raw}
    \bm y_{k+1}
    =
    \begin{pmatrix}
        \bm y_k - \alpha\, \bm a_k\\
        \alpha
    \end{pmatrix},
    \quad
    \alpha=\frac{x_{k+1}-\bm r_k^\top \bm y_k}{s_k}.
\end{equation}

To update $\bm y_{k+1}$ by \eqref{eq:y-update-raw} for $1 \leq k \leq n-1$,
we need $(\bm a_k,s_k)$ for each $k$, hence we derive their recursions.
Let $J_k$ be the exchange matrix of order $k$: 
\begin{equation}
    (J_k)_{ij}= 
    \begin{cases}
        1\quad \mathrm{if} \quad i+j=k+1, \\
        0 \quad \mathrm{otherwise},
    \end{cases}
\end{equation}
such that $J_k(v_1,\dots,v_k)^\top=(v_k,\dots,v_1)^\top$. According to Eq.~\eqref{eq:def_ak}, we have
\begin{equation}
    \label{eq:ak+1}
    \begin{pmatrix}
        c_0 & (J_k \bm r_k)^{\top} \\
        J_k \bm r_k & C_k
    \end{pmatrix}
    \begin{pmatrix}
        \kappa \\
        \bm p
    \end{pmatrix}
    = 
    \begin{pmatrix}
        c_{k+1} \\
        \bm r_k 
    \end{pmatrix}
    ,
\end{equation}
with 
\begin{equation}
    \kappa \coloneqq a_{k+1}^{(1)}, 
\end{equation}    
and 
\begin{equation}
    \bm p \coloneqq (a_{k+1}^{(2)}, a_{k+1}^{(3)},\dots, a_{k+1}^{(k+1)})^\top, 
\end{equation}
Equation~\eqref{eq:ak+1} implies that
\begin{equation}
    \bm p = \bm a_k - \kappa C_k^{-1} J_k \bm r_k,
\end{equation}
and
\begin{equation}
    \kappa
    =
    \frac{c_{k+1}-\bm r_k^\top J_k \bm a_k}{c_0 - \bm r_k^\top J_k C_k^{-1} J_k \bm r_k}.
\end{equation}
Since the centrosymmetricity of $C_k^{-1}$, i.e.,  
\begin{equation}
    J_k C_k^{-1} = C_k^{-1} J_k,
\end{equation} 
follows from that of the symmetric Toeplitz matrix $C_k$, we obtain
\begin{equation}
    \label{eq:a-update}
    \bm a_{k+1}
    =
    \begin{pmatrix}
        \kappa \\
        \bm a_k - \kappa\, J_k \bm a_k\\
    \end{pmatrix},
    \quad
    \kappa=\frac{c_{k+1}-\bm r_k^\top J_k \bm a_k}{s_k}.
\end{equation}
Finally, we have
\begin{align}
    \label{eq:s-update}
    s_{k+1} &= c_0 - \bm r_{k+1}^\top \bm a_{k+1} \notag \\
    &= c_0 - (c_{k+1}, \bm r_k^\top) 
    \begin{pmatrix}
        \kappa \\
        \bm a_k - \kappa\, J_k \bm a_k\\
    \end{pmatrix} \notag \\
    &= s_k (1 - \kappa^2). 
\end{align}

In summary, we can compute the solution $\bm y_n = C_n^{-1} \bm x_n$ recursively by Eqs.~\eqref{eq:y-update-raw}, \eqref{eq:a-update}, and \eqref{eq:s-update} with initial conditions 
\begin{gather}
    \bm y_1=\frac{x_1}{c_0},
    \quad
    \bm a_1=\frac{c_1}{c_0},
    \quad \mathrm{and} \quad 
    s_1=c_0-\frac{c_1^2}{c_0}. 
\end{gather}
The determinant of $C_n$ can likewise be derived recursively by a Schur complement formula~\cite{serre2010}: 
\begin{align}
    \det C_{k+1} = 
    \left|
    \begin{matrix}
        \huge{C_k} & \bm{r}_{k} \\
        \bm{r}_{k}^{\top} & c_0 
    \end{matrix}
    \right|
    = \det C_k s_k
    ,
\end{align}
and hence
\begin{align}
    \det C_{n} 
    = c_0 \prod_{k=1}^{n-1} s_k
    .
\end{align}

\clearpage
\section{Details in numerical integration with importance sampling}
\label{SI:sec:detail_IS}
We determine the Gaussian-mixture centers $\{\tilde{\bm{\theta}}_j\}$ by maximizing $\rho(\bm \theta)$ using the L-BFGS quasi-Newton algorithm in Julia~\cite{optim}. We perform the optimization in log-parameter space. Specifically, we set $\phi_i = \ln \theta_i$, and maximize $\kappa(\bm \phi) \coloneqq \rho((e^{\phi_1},\ldots,e^{\phi_\nu})^{\top})$ with $\bm \phi \coloneqq (\phi_1, \ldots, \phi_{\nu})^{\top}$. Multiple local maxima are detected by repeating the optimization from $100$ different initial conditions generated by a Sobol sequence~\cite{sobol1967distribution} over the prior domain. The repeated searches are terminated once $20$ consecutive initializations converge to an already detected local maximum, reducing unnecessary numerical cost. Using the obtained modes $\{\tilde{\bm \theta}_j \}$, we set the GMM weights $\pi_j$ in Eq.~\eqrefstar{eq:proposal_gmm} by applying a tempered softmax to the log unnormalized posterior values at the modes. Specifically, with
\begin{equation}
    R_j \coloneqq \rho(\tilde{\bm \theta}_j),
    \qquad
    R_{\max} \coloneqq \max_{1\leq j\leq K} R_j,
\end{equation}
we use
\begin{equation}
    \pi_j \coloneqq
    \frac{\exp\left[(R_j-R_{\max})/z\right]}
    {\sum_{i=1}^{K}\exp\left[(R_i-R_{\max})/z\right]},
\end{equation}
with a positive parameter $z$, which we set $z=2.0$ by monitoring the ESS.

For each mode, we next construct the covariance matrix used in the corresponding proposal component. When the local curvature is well behaved, the covariance is obtained from a local Laplace approximation of $\rho(\bm \theta)$ around the mode $\tilde{\bm{\theta}}_j$. Specifically, we compute the Hessian
\begin{equation}
    \mathcal{H}_j \coloneqq \left. \nabla_{\bm \theta}^2 \rho(\bm \theta) \right|_{\bm \theta = \tilde{\bm \theta}_j},
\end{equation}
and, if $\mathcal{H}_j$ is negative definite, set
\begin{equation}
    \label{eq:cov_from_hessian}
    \Lambda_j = - c^2 \, \mathcal{H}_j^{-1},
\end{equation}
where $c>1$ is an inflation factor that gives the proposal broader tails than the local Laplace approximation; in the present implementation, we use $c=4.0$.
In some cases, $\mathcal{H}_j$ contains nonfinite entries or is not negative definite, so the covariance cannot be constructed from the inverse negative Hessian. This can occur when $\tilde{\bm{\theta}}_j$ lies near the boundary of the prior support or when the local curvature is poorly resolved numerically. In such cases, we retain the mode $\tilde{\bm{\theta}}_j$ and use a diagonal fallback covariance whose scale is set relative to the mode value of each parameter. For the $\ell$-th parameter $\theta^{(\ell)}$ with mode value $\tilde{\theta}_{j}^{(\ell)}$, we define
\begin{equation}
\label{eq:fb_from_hessian}
    \Lambda_j^{\mathrm{fb}}
    =
    c_f^2 \, \mathrm{diag}\!\left(
    \left[\tilde{\theta}_{j}^{(1)}\right]^2,\ldots,\left[\tilde{\theta}_{j}^{(\nu)}\right]^2
    \right),
\end{equation}
where $c_f$ is a positive parameter and $\nu$ denotes the number of model parameters (i.e., the dimension of $\bm \theta$).
Because the priors are log-uniform, parameter uncertainty is more naturally measured relative to the mode value than the prior width. We therefore set each fallback standard deviation proportional to the corresponding mode value, using $c_f=2.0$. 

When using the fallback covariance $\Lambda_j^{\mathrm{fb}}$~\eqref{eq:fb_from_hessian}, we use the multivariate Gaussian proposal $\mathcal{N}(\bm \theta; \tilde{\bm \theta}_j, \Lambda_j^{\mathrm{fb}})$ truncated to the prior support. This prevents proposal mass from leaking outside the prior domain, especially when the mode $\tilde{\bm \theta}_j$ is close to the prior boundary. When using the Hessian-based covariance matrix $\Lambda_j$~\eqref{eq:cov_from_hessian}, we keep the untruncated multivariate Gaussian proposal $\mathcal{N}(\bm \theta; \tilde{\bm \theta}_j, \Lambda_j)$.

\clearpage
\section{Evaluation of parameter estimation}
\label{SI:sec:eval_para_est}
\begin{figure}[h]
    \centering
    \includegraphics[width=1.0
    \linewidth]{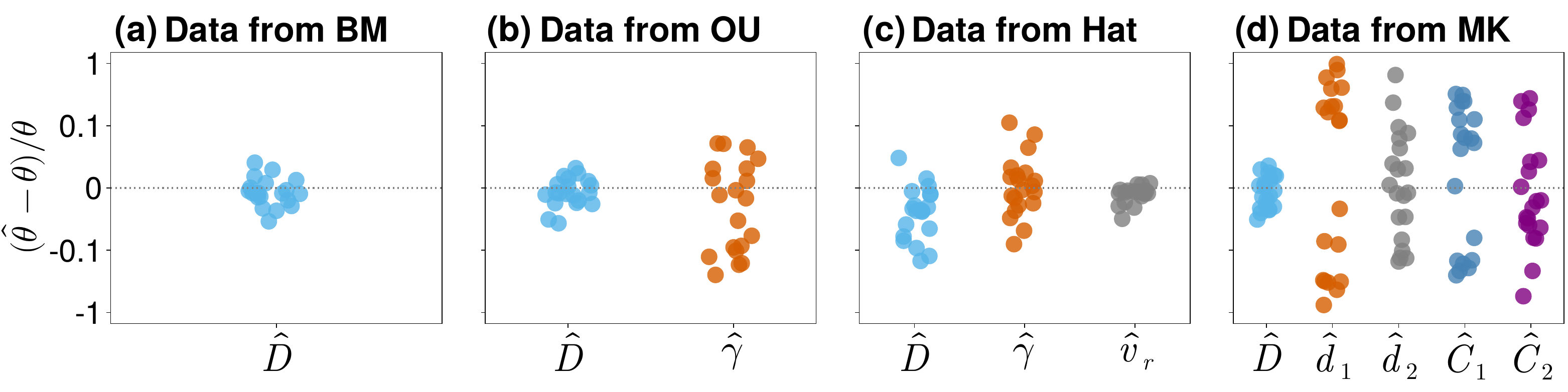}
    \caption{The relative error of parameter estimation from synthetic trajectories. We use the same trajectories and priors as those in Fig.~\ref*{fig:verify_framework_summary} in the main article. Panels (a), (b), (c), and (d) correspond to the results for trajectories generated from the BM, OU, Mexican-hat, and memory kernel models, respectively. 
    }
    \label{fig:wide_prior_paraest}
\end{figure}

Using the same synthetic-data as in Fig.~\ref*{fig:verify_framework_summary} of the main text, we additionally evaluate the accuracy of parameter estimation for the model that matches the data-generating model. Figure~\ref{fig:wide_prior_paraest} shows the relative errors of the corresponding MAP estimates. As shown in panels (a), (b), and (c), the relative errors for the BM, OU, and Mexican-hat models are typically about $10\%$, apart from a few outliers. The errors are larger for the memory-kernel model [Fig.~\ref{fig:wide_prior_paraest}~(d)], partly because estimating the smaller damping rate $d_1$ is challenging. This difficulty is expected for finite-length trajectories: $d_1$ controls the long-lag behavior of the VACF [Eq.
~\eqrefstar{eq:fly_VACF}], whereas empirical estimates at large lag $\tau$ become noisy because they are averaged over fewer trajectory segments.

\clearpage
\section{Numerical results with narrow priors}
\label{SI:app:narrow_prior}
\begin{figure}[h]
    \centering
    \includegraphics[width=0.6\linewidth]{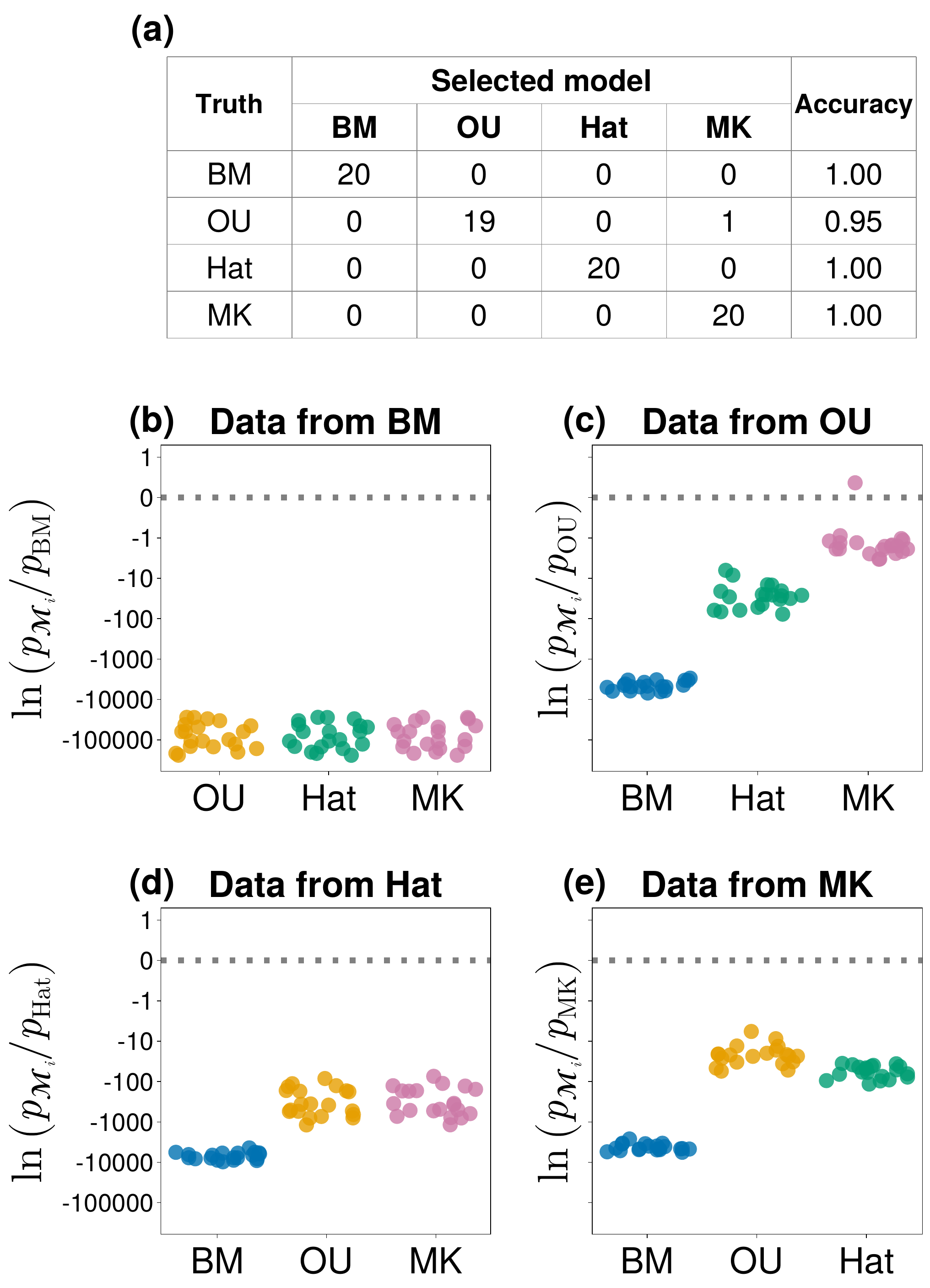}
    \caption{Same analysis as in Fig.~\ref*{fig:verify_framework_summary} of the main text, however with narrower priors. For all parameters in the BM and OU models, for $D$ and $\gamma$ in the Mexican-hat model, and for $D$ and $d_1$ in the memory-kernel model, we use $\mathcal{L}(10^{-2}, 10.0)$ as the prior distribution. The remaining parameters, namely $v_r$ in the Mexican-hat model and $u$ and $s$ in the memory-kernel model, use the same priors as in Table~\ref*{tab:prior} of the main text. We use the same synthetic data as in Fig.~\ref*{fig:verify_framework_summary}, i.e., $20$ trajectories with $T=200$ and $\Delta t = 0.1$ generated from each of the four candidate models. }
    \label{fig:narrow_prior}
\end{figure}

To investigate how model comparison depends on prior width, we repeat the numerical experiment in Fig.~\ref*{fig:verify_framework_summary} of the main text using narrower priors for all parameters except the model-specific parameters, namely $v_r$ in the Mexican-hat model and $u$ and $s$ in the memory kernel model. The results are summarized in Fig.~\ref{fig:narrow_prior}, and the priors used are described in the figure caption. Figure~\ref{fig:narrow_prior} (a) shows the same results as those obtained with the original priors (Fig.~\ref*{fig:verify_framework_summary}): the framework reliably selects the BM, Mexican-hat, and memory-kernel models. The relatively rare misclassifications occur for data generated from the OU model, and these cases are accompanied by log Bayes factors close to zero [Fig.~\ref{fig:narrow_prior}~(c)]. 

\clearpage
\section{Sampling-interval dependence of log Bayes factors using data from other models}
\label{SI:sec:dif_dt_other}

\begin{figure}[h]
    \centering
    \begin{minipage}{0.48\linewidth}
        \centering
        \includegraphics[width=\linewidth]{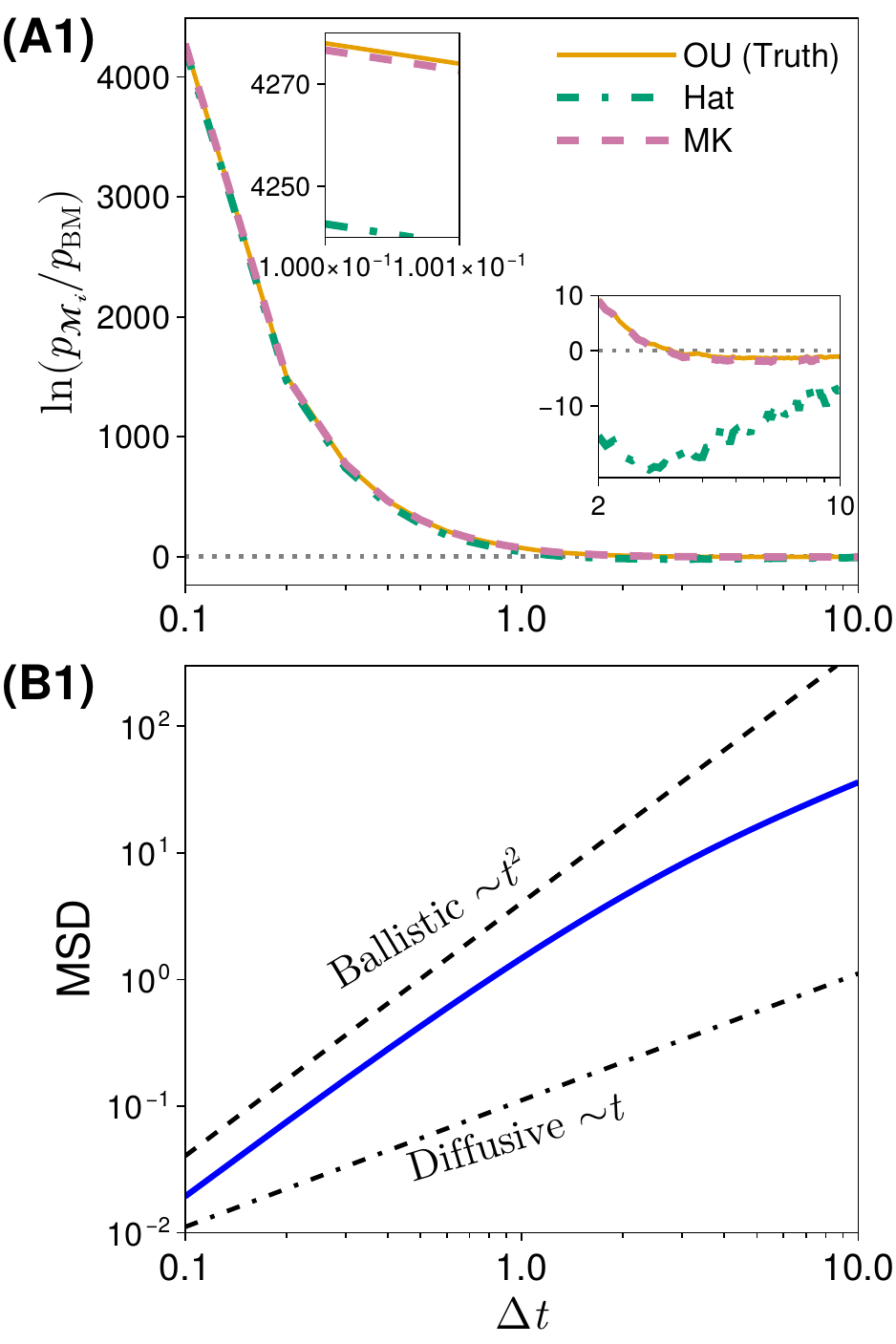}
    \end{minipage}
    \hfill
    \begin{minipage}{0.48\linewidth}
        \centering
        \includegraphics[width=\linewidth]{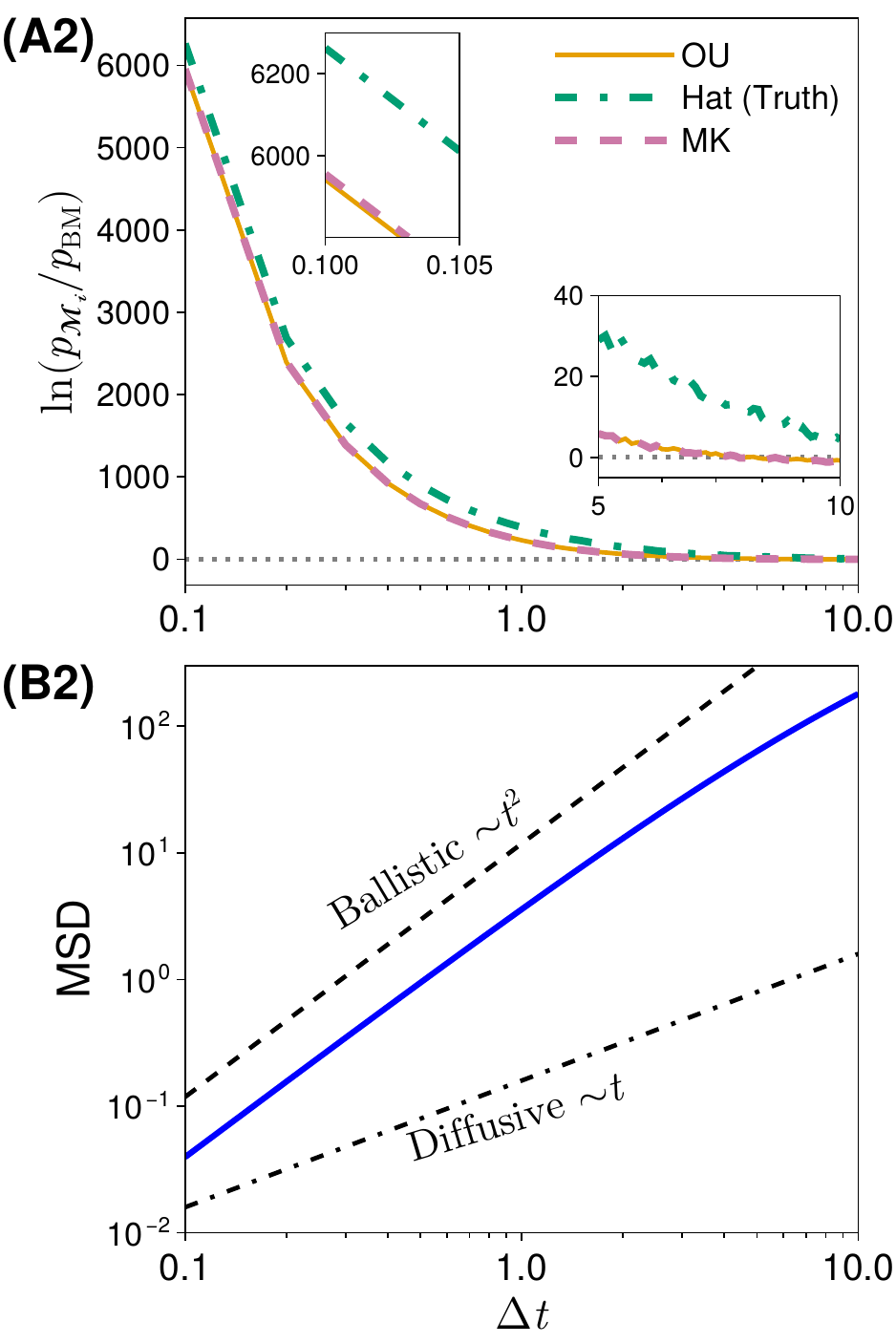}
    \end{minipage}
    \caption{Same analysis as in Fig.~\ref*{fig:comp_artificial_meanstd}, however using data generated from the OU model [left column; panels (A1) and (B1)] and the Mexican-hat model [right column; panels (A2) and (B2)]. 
    The original trajectories are generated with $T=200$ and $\Delta t=0.01$, and trajectories at coarser sampling intervals are obtained by subsampling. The parameters are fixed at $\gamma=1.0$ and $D=1.0$ for the OU model; $\gamma=1.0$, $v_r=2.0$, and $D=1.0$ for the Mexican-hat model. For each generating model, we use $20$ trajectories generated with different random seeds. The figure conventions are the same as those in Fig.~\ref*{fig:comp_artificial_meanstd}. }
    \label{fig:comp_artificial_meanstd_other}
\end{figure}

Figure~\ref{fig:comp_artificial_meanstd_other} shows the dependence of the log Bayes factors and displacement statistics on the sampling interval $\Delta t$ for trajectories generated from the OU and Mexican-hat models. For the OU model, the theoretical MSD is evaluated analytically using Eq.~\eqrefstar{eq:OU_msd} with the same parameter values as those used to generate the trajectories. For the Mexican-hat model, for which an analytical MSD is unavailable, we generate the same number of independent trajectories using the same parameter values but different noise realizations and plot the ensemble-averaged TAMSD $\langle \delta(\tau) \rangle$. As in the case when trajectories are generated from the memory kernel model [Fig.~\ref*{fig:comp_artificial_meanstd} in the main text], the true generating model is generally favored at fine sampling intervals, whereas the log Bayes factors become less decisive as $\Delta t$ increases. The corresponding MSDs also shift from the short-time ballistic regime toward the long-time diffusive regime. These results confirm that the loss of model identifiability under temporal coarse-graining is not specific to synthetic trajectories generated from the memory kernel model, but is observed consistently across the second-order models considered here.


\clearpage
\section{Comparison between exact and approximate likelihoods for the OU model}
\label{app:exact_approx_OU}
\begin{figure}[h]
    \centering
    \includegraphics[width=.5\linewidth]{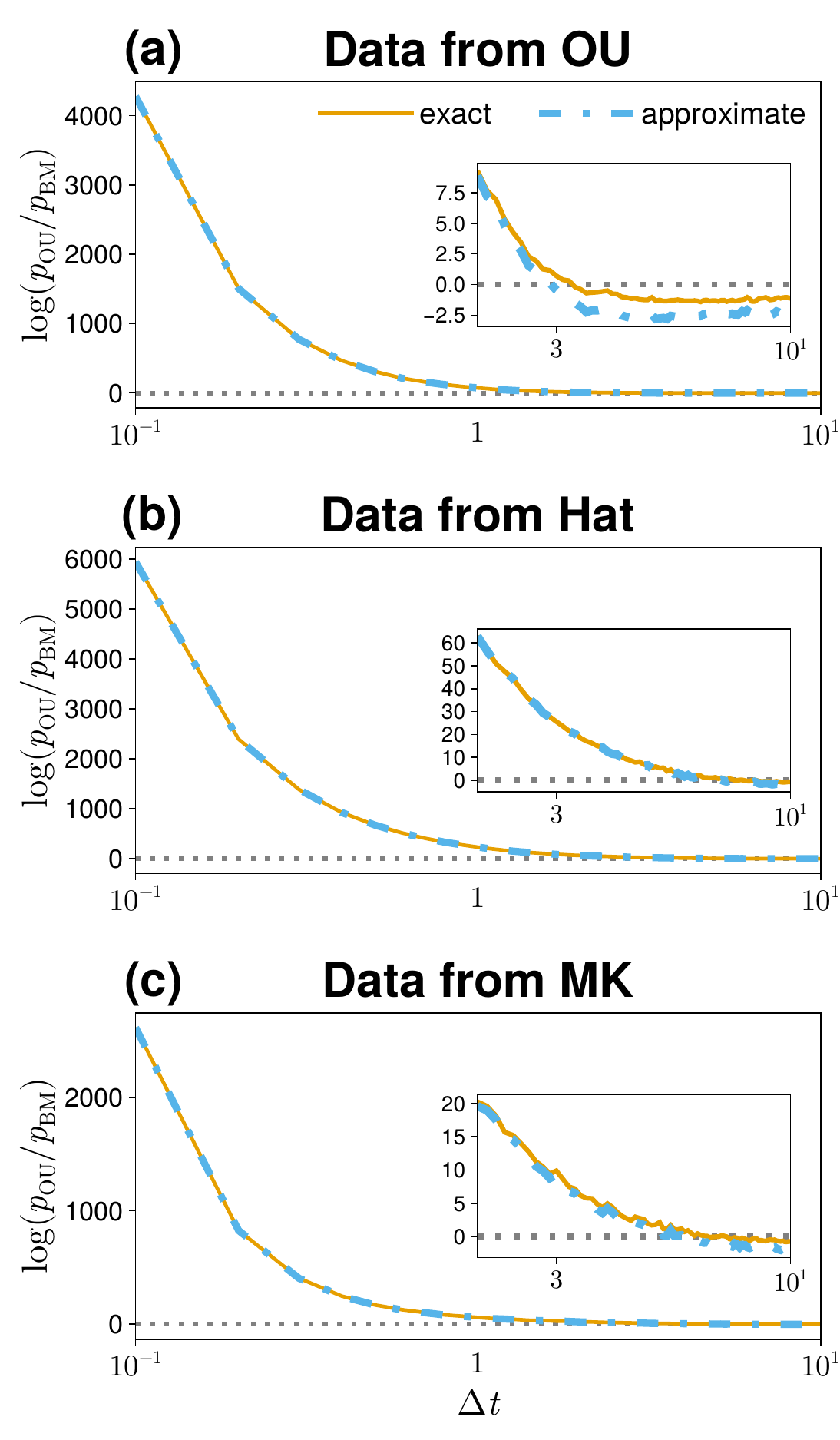}
    \caption{Comparison of log Bayes factors computed using the exact OU likelihood (yellow solid lines) and the approximate OU likelihood (blue dash-dotted lines). The synthetic trajectories in panels (a), (b), and (c) are generated from the OU, Mexican-hat, and memory kernel models, respectively, using the same parameter values, trajectory ensembles, and subsampling procedure as in Figs.~\ref*{fig:comp_artificial_meanstd} and~\ref*{fig:comp_artificial_meanstd_other}. Each panel shows the log Bayes factor relative to the BM model as a function of the sampling interval $\Delta t$. Insets provide enlarged views of the results at large sampling intervals.}
    \label{fig:OUvsOUexact}
\end{figure}

To assess the validity of the approximate likelihood [i.e., Eqs.~\eqrefstar{eq:likeli_nonlin_secant} and \eqrefstar{eq:likeli_nonlin_approx}] in marginal-likelihood calculations at different sampling intervals $\Delta t$, we apply the same approximation to the linear OU model, for which the exact likelihood is available. We then compare the resulting marginal likelihoods with those obtained from the exact OU likelihood. Figure~\ref{fig:OUvsOUexact} summarizes the results using the same synthetic trajectories as in Fig.~\ref*{fig:comp_artificial_meanstd} and Fig.~\ref{fig:comp_artificial_meanstd_other}. In panels (a), (b), and (c), where the trajectories are generated from the OU, Mexican-hat, and memory kernel models, respectively, the log Bayes factors obtained from the approximate and exact OU likelihoods agree well over a wide range of $\Delta t$.

\clearpage
\section{Model comparison results of other strains}
\label{app:other_strain}
\begin{figure}[h]
    \centering
    \includegraphics[width=.5\linewidth]{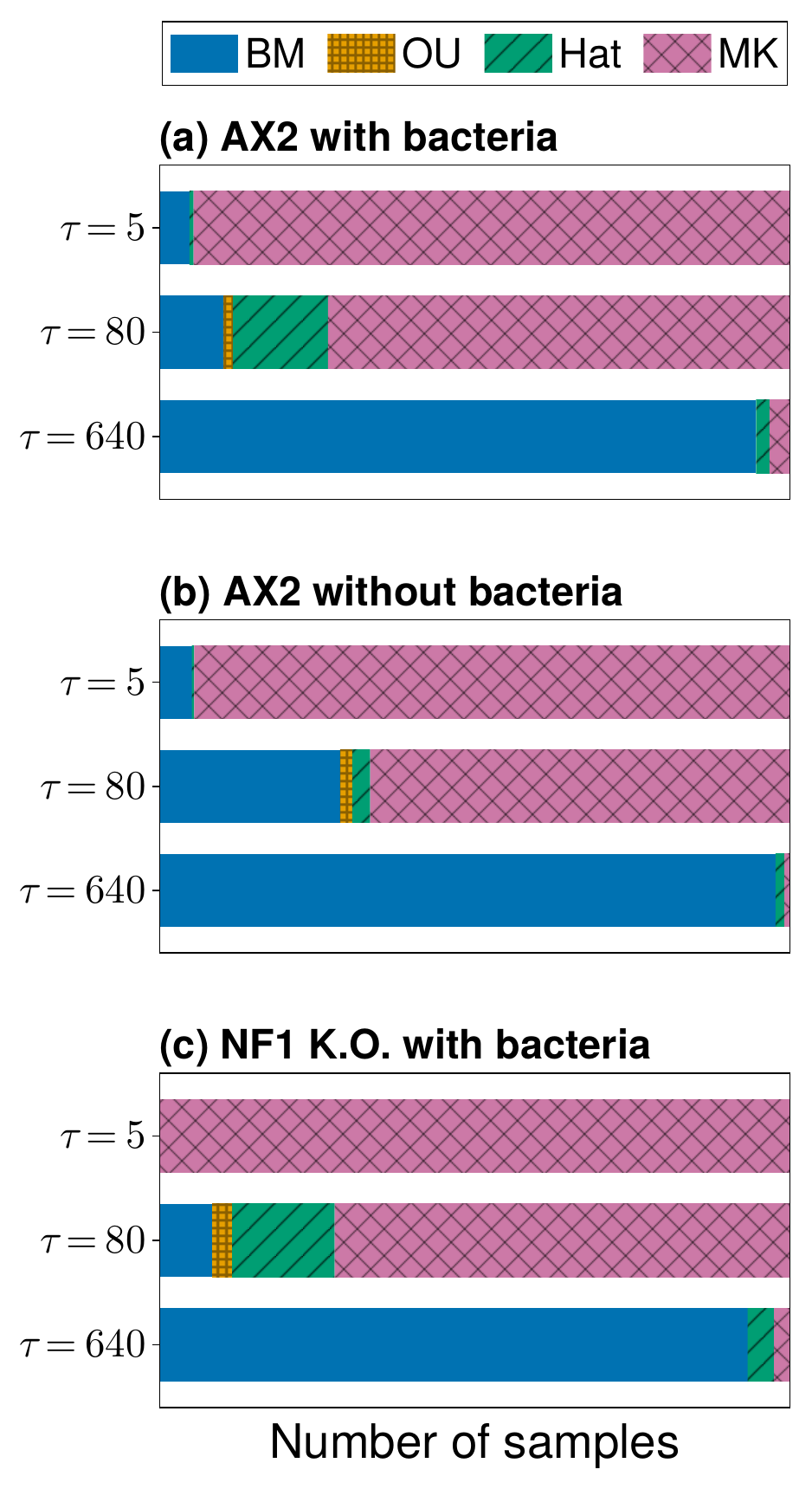}
    \caption{Counts of selected models across sampling intervals for trajectories from (a) AX2 cells with bacteria, (b) AX2 cells without bacteria, and (c) NF1 knockout cells with bacteria. Figure conventions are the same as in Fig.~\ref*{fig:DdB_comp_summary} of the main text. Stacked bars indicate the number of trajectories assigned to each model; the total number of trajectories differs among experimental conditions, although the horizontal width is kept the same across panels. 
    }
    \label{fig:other_comp_summary}
\end{figure}

\begin{figure}[h]
    \centering
    \includegraphics[width=.55\linewidth]{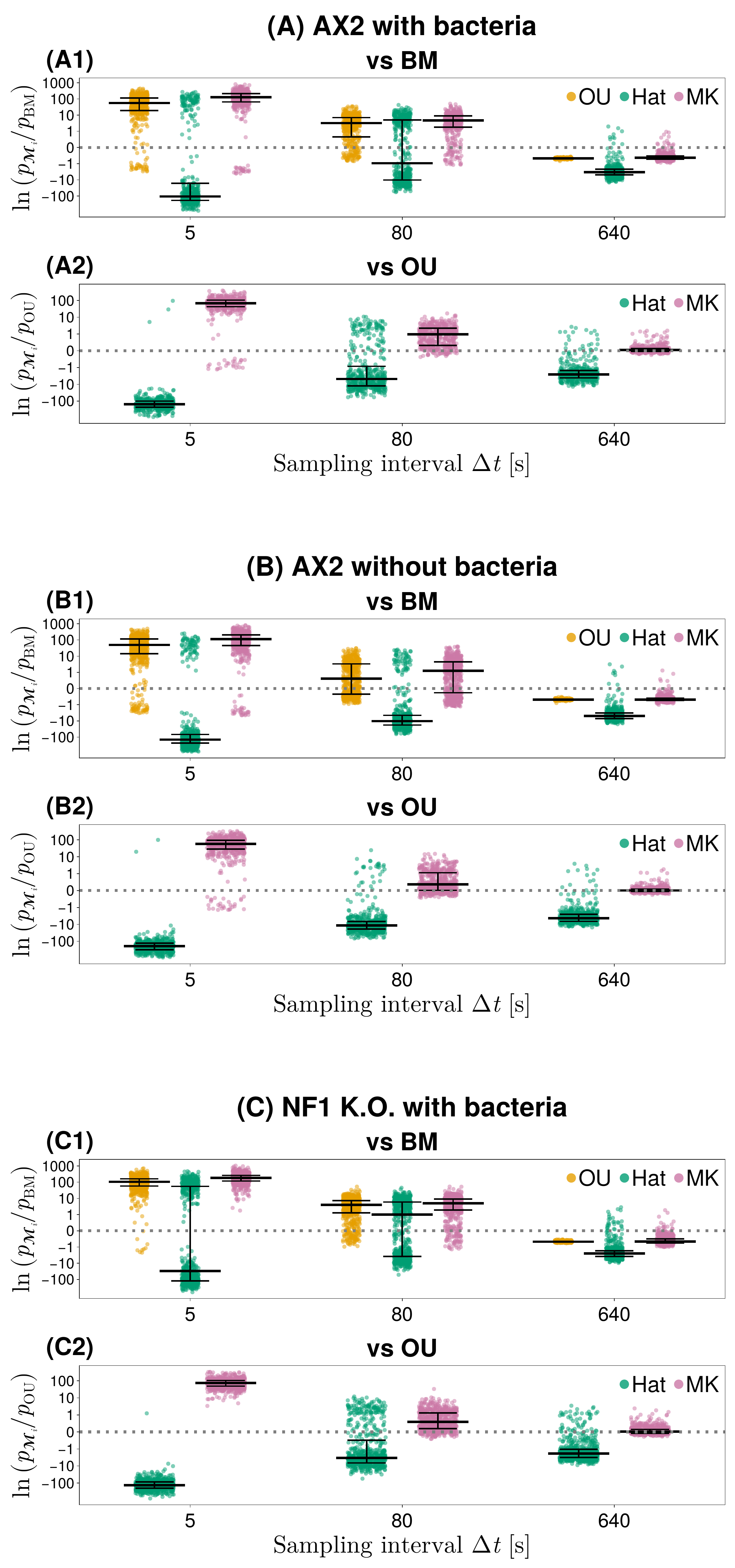}
    \caption{Scatter plots of log Bayes factors at different sampling intervals for AX2 cells with bacteria [panels (A1) and (A2)], AX2 cells without bacteria [panels (B1) and (B2)], and NF1 knockout cells with bacteria [panels (C1) and (C2)]. Panels (A1), (B1), and (C1) denote the log Bayes factors relative to the BM model, whereas panels (A2), (B2), and (C2) illustrate those relative to the OU model. Other figure conventions are the same as in Fig.~\ref*{fig:DdB_comp_vsBM_vsOU} of the main text. 
    }
    \label{fig:other_comp_vsBM_vsOU}
\end{figure}

To assess whether the effect of temporal coarse-graining observed for the DdB trajectories is dataset-specific or reflects a more general feature of position-only model comparison, we repeat the analysis for additional strains and experimental conditions: AX2 cells with bacteria, AX2 cells without bacteria, and NF1 knockout cells with bacteria. Figure~\ref{fig:other_comp_summary} summarizes the model assignments, showing the same tendency as in the DdB trajectories (Fig.~\ref*{fig:DdB_comp_summary} of the main text): coarser sampling intervals lead to more frequent selection of simpler models. We further examine the distributions of log Bayes factors relative to the BM and OU models in Fig.~\ref{fig:other_comp_vsBM_vsOU}. Across all strains, the qualitative trend is consistent with the DdB results: at short sampling intervals, the log Bayes factors more clearly separate competing models, whereas at longer sampling intervals the evidence differences contract and multiple candidate models become difficult to distinguish. 

\clearpage
\section{Effective sample sizes for synthetic and experimental trajectories}
\label{app:ess}
\begin{figure}[h]
    \centering
    \includegraphics[width=.9\textwidth]{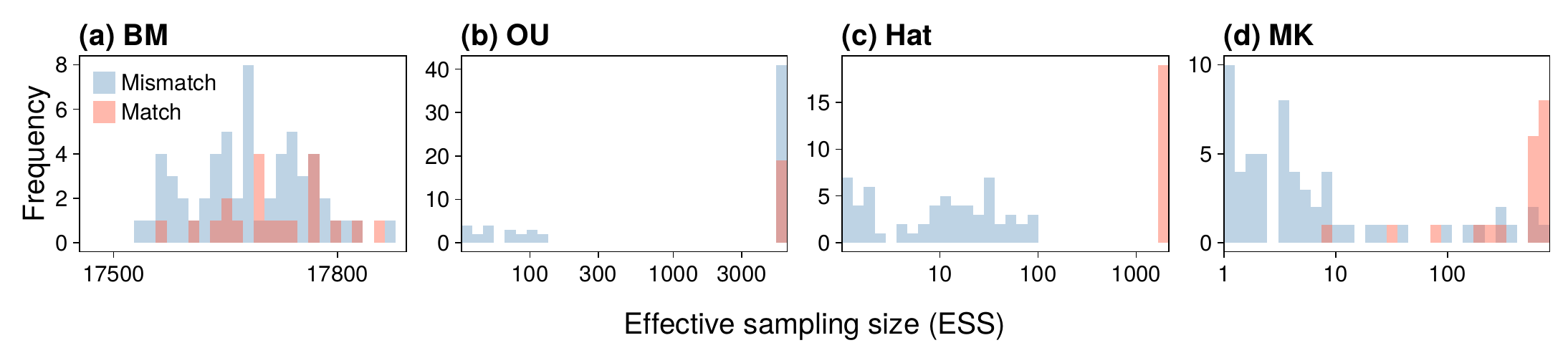}
    \caption{Histograms of ESS values calculated for the synthetic-data experiment in Fig.~\ref*{fig:verify_framework_summary}. Panels (a), (b), (c), and (d) correspond to the marginal likelihood estimation for the BM, OU, Mexican-hat, and memory kernel models, respectively. In each panel, the histogram shows ESS values for all $80$ synthetic trajectories. Red bars indicate cases in which the candidate model matches the data-generating model, whereas blue bars indicate model-mismatch cases. }
    \label{fig:ESS_hist_synthetic}
\end{figure}

\begin{figure}[h]
    \centering
    \includegraphics[width=.9\textwidth]{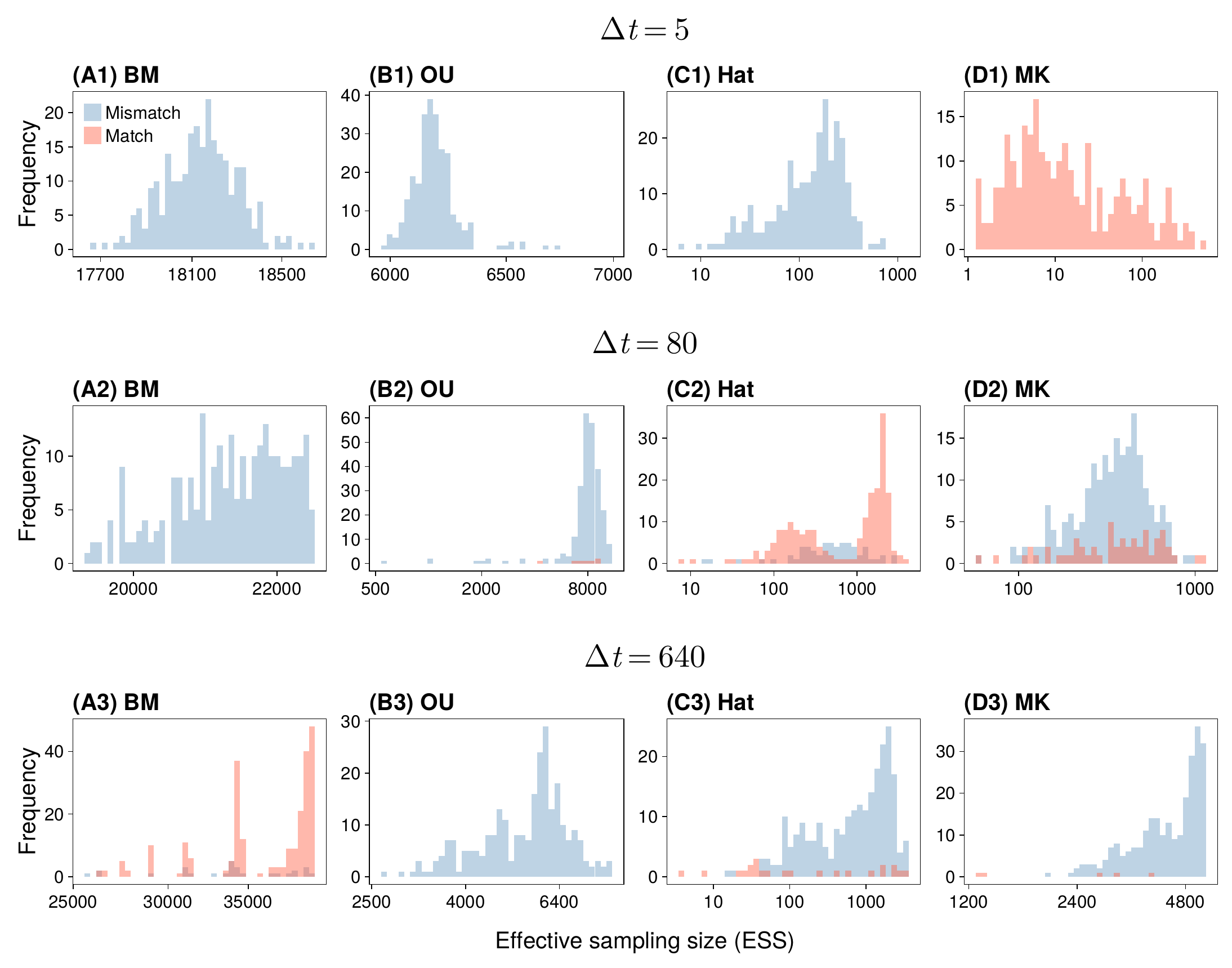}
    \caption{Histograms of ESS values calculated for model comparison of the DdB trajectories. The top, middle, and bottom rows correspond to sampling intervals $\Delta t = 5$, $80$, and $\qty{640}{\second}$, respectively. The first, second, third, and fourth columns correspond to marginal-likelihood estimates for the BM, OU, Mexican-hat, and memory kernel models, respectively. Each panel shows the ESS distribution across all $261$ DdB trajectories. Red bars indicate trajectories for which the corresponding candidate model has the largest marginal likelihood, whereas blue bars indicate trajectories for which another model is selected. }
    \label{fig:ESS_hist_DdB}
\end{figure}

Figures~\ref{fig:ESS_hist_synthetic} and~\ref{fig:ESS_hist_DdB} show histograms of the ESS values defined in Eq.~\eqrefstar{eq:ess}. Figure~\ref{fig:ESS_hist_synthetic} corresponds to the synthetic-data experiment in Fig.~\ref*{fig:verify_framework_summary} in the main text. Red bars indicate model-match cases, in which the candidate model coincides with the data-generating model, whereas blue bars indicate model-mismatch cases. Figure~\ref{fig:ESS_hist_DdB} shows the corresponding ESS values for the DdB trajectories. In this case, red bars indicate trajectories for which the corresponding candidate model has the largest marginal likelihood, whereas blue bars indicate trajectories for which another model is selected.

For the BM and OU models, the ESS values are generally high ($\gg 1000$). For the more complex Mexican-hat and memory-kernel models, the ESS can be as low as $\sim 10$, especially in model-mismatch cases. In most cases, however, the ESS remains larger, typically $\gtrsim 100$.

\clearpage
\bibliography{motility.bib}